\documentclass[fleqn,usenatbib]{mnras}

\usepackage{lineno}
\usepackage[dvipsnames]{xcolor}

\usepackage[T1]{fontenc}

\DeclareRobustCommand{\VAN}[3]{#2}
\let\VANthebibliography\thebibliography
\def\thebibliography{\DeclareRobustCommand{\VAN}[3]{##3}\VANthebibliography}

\usepackage{graphicx}	
\usepackage{amsmath}	
\usepackage{amssymb}	
\usepackage{bm}

\newcommand{\angstrom}{\textup{\AA}}
\newcommand{\hubble}{\ensuremath{\mathrm{~km} \mathrm{~s}^{-1} \mathrm{Mpc}^{-1}}}

\title[]{Calibrating the Absolute Magnitude of Type Ia Supernovae in Nearby Galaxies using [OII] and Implications for $H_{0}$}

\author[Dixon et al.]{
\parbox{\textwidth}{\Large
M.~Dixon,$^{1}$
J.~Mould,$^{1}$
C.~Lidman,$^{2,3}$
E.~N.~Taylor,$^{1}$
C.~Flynn,$^{1}$
A.~R.~Duffy,$^{1}$
L.~Galbany,$^{4,5}$
D.~Scolnic,$^{6}$
T.~M.~Davis,$^{7}$
A.~M\"oller,$^{1}$
L.~Kelsey,$^{8,9}$
J.~Lee,$^{10}$
P.~Wiseman,$^{9}$
M.~Vincenzi,$^{8,9}$
P.~Shah,$^{11}$
M.~Aguena,$^{12}$
S.~S.~Allam,$^{13}$
O.~Alves,$^{14}$
D.~Bacon,$^{8}$
S.~Bocquet,$^{15}$
D.~Brooks,$^{11}$
D.~L.~Burke,$^{16,17}$
A.~Carnero~Rosell,$^{12,18}$
D.~Carollo,$^{51}$
J.~Carretero,$^{19}$
C.~Conselice,$^{20,21}$
L.~N.~da Costa,$^{12}$
M.~E.~S.~Pereira,$^{22}$
H.~T.~Diehl,$^{13}$
P.~Doel,$^{11}$
S.~Everett,$^{23}$
I.~Ferrero,$^{24}$
B.~Flaugher,$^{13}$
J.~Frieman,$^{13,25}$
J.~Garc\'ia-Bellido,$^{26}$
M.~Gatti,$^{10}$
E.~Gaztanaga,$^{4,5,8}$
G.~Giannini,$^{19,25}$
D.~Gruen,$^{15}$
R.~A.~Gruendl,$^{27,28}$
G.~Gutierrez,$^{13}$
K.~Herner,$^{13}$
S.~R.~Hinton,$^{7}$
D.~L.~Hollowood,$^{29}$
K.~Honscheid,$^{30,31}$
D.~J.~James,$^{32}$
K.~Kuehn,$^{33,34}$
M.~Lima,$^{12,35}$
J.~L.~Marshall,$^{36}$
J. Mena-Fern{\'a}ndez,$^{37}$
F.~Menanteau,$^{27,28}$
R.~Miquel,$^{19,38}$
J.~Myles,$^{39}$
R.~C.~Nichol,$^{40}$
R.~L.~C.~Ogando,$^{41}$
A.~Palmese,$^{42}$
A.~Pieres,$^{12,41}$
A.~A.~Plazas~Malag\'on,$^{16,17}$
S.~Samuroff,$^{43}$
E.~Sanchez,$^{44}$
D.~Sanchez Cid,$^{44}$
I.~Sevilla-Noarbe,$^{44}$
M.~Smith,$^{9}$
F.~Sobreira,$^{12,45}$
E.~Suchyta,$^{46}$
M.~E.~C.~Swanson,$^{27}$
G.~Tarle,$^{14}$
C.~To,$^{30}$
B.~E.~Tucker,$^{3}$
D.~L.~Tucker,$^{13}$
V.~Vikram,$^{47}$
A.~R.~Walker,$^{48}$
and N.~Weaverdyck$^{49,50}$
\begin{center} (DES Collaboration) \end{center}
}}

\usepackage{eso-pic}

\date{Accepted XXX. Received YYY; in original form ZZZ}

\pubyear{2024}

\begin{document}
\label{firstpage}
\pagerange{\pageref{firstpage}--\pageref{lastpage}}
\maketitle

\begin{abstract}
The present state of cosmology is facing a crisis where there is a fundamental disagreement in measurements of the Hubble constant ($H_{0}$), with significant tension between the early and late universe methods. Type Ia supernovae (SNe Ia) are important to measuring $H_{0}$ through the astronomical distance ladder. However, there remains potential to better standardise SN Ia light curves by using known dependencies on host galaxy properties after the standard light curve width and colour corrections have been applied to the peak SN Ia luminosities. To explore this, we use the 5-year photometrically identified SNe Ia sample obtained by the Dark Energy Survey, along with host galaxy spectra obtained by the Australian Dark Energy Survey. Using host galaxy spectroscopy, we find a significant trend with the equivalent width (EW) of the [OII] $\lambda\lambda$ 3727, 29 doublet, a proxy for specific star formation rate, and Hubble residuals. We find that the correlation with [OII] EW is a powerful alternative to the commonly used mass step after initial light curve corrections. Applying this [OII] EW correction to 20 SNe Ia in calibrator galaxies observed with WiFeS, we examined the impact on SN Ia absolute magnitudes and $H_{0}$. Our [OII] EW corrections result in $H_{0}$ values ranging between 73.04 to 73.51$\hubble$, with a combined statistical and systematic uncertainty of $\sim$1.31$\hubble$. However, even with this additional correction, the impact of host galaxy properties in standardising SNe Ia appears limited in reducing the current tension ($\sim$5$\sigma$) with the CMB result for $H_{0}$.
\end{abstract}


\begin{keywords}
cosmology: observations, galaxies: general,  transients: supernovae
\end{keywords}



\section{Introduction}
Cosmology is currently facing a crisis known as the Hubble tension, where there is a disagreement between different methods measuring the Hubble constant ($H_{0}$). The most significant discrepancy is between the values derived from the cosmic microwave background (CMB) \citep{Planck2018} and the local distance ladder \citep{Riess_2022}. Currently, the Hubble tension is at a significance level of 5$-$6$\sigma$ \citep{Di_Valentino_2021}. 

The local and direct approach to measuring $H_{0}$ is through the astronomical distance ladder, which relies on the use of standardisable candles, such as Type Ia supernovae (SNe Ia). SNe Ia are excellent cosmological probes in understanding the expansion rate of the universe. By calibrating the absolute magnitudes of SNe Ia ($M^{B}_{0}$) in nearby galaxies, one can use SNe Ia in the Hubble flow and measure $H_{0}$ \citep{Dhawan_2020, Khetan2020, Freedman_2021, Riess_2022}. Using 42 SNe Ia for which host galaxy distances are available using Cepheids, \cite{Riess_2022} found $M^{B}_{0} = -19.253 \pm 0.027$ mag, and determined $H_{0} = 73.04 \pm 1.04 \hubble$.

By increasing the sample of nearby SNe Ia in galaxies that have distance estimates from Cepheids, one can increase the precision in the measurement of $H_{0}$. However, the current rate of finding nearby galaxies with SN Ia and Cepheid distances is $\sim$1/year \citep{Riess_2022}. Alternatively, the sample of SN Ia calibrator galaxies can be increased by including hosts that have distances derived using the tip of the red giant branch (TRGB; \citealt{Freedman_2021}) or surface brightness fluctuations (SBF; \citealt{Jensen_2021}). In addition to reducing statistical uncertainties, these approaches will help better understand and control systematic uncertainties.

To standardise SNe Ia for use in cosmology, corrections are made to account for the relationship between the peak magnitude and the width of SN Ia light curve \citep{1993ApJ...413L.105P} and the SN Ia colour \citep{1998A&A...331..815T}. To further reduce the intrinsic scatter, many studies have used broad-band photometry and spectroscopy of the host galaxies to correct for trends between the colour and width-corrected luminosities of SNe Ia and the properties of their host galaxies. Correlations with Hubble residuals have been found with host gas phase metallicity \citep{2011d'andrea,Pan2013, Moreno-Raya_2018}, stellar age \citep{Childress_2013,2019rose}, specific star formation rate \citep{Lampeitl_2010, 2011d'andrea, Childress_2013, Rigault_2015, Rigault_2020, Briday_2022, Dixon_2022, Galbany_2022, Martin_2024},  rest-frame colour \citep{2018roman, 2021Kelsey, Kelsey_2023} and host galaxy dust \citep{brout2021dust, Dixon_2022, meldorf2022}.

The most common correction is with host galaxy stellar mass (\citealp{kelly2010,Sullivan_2010,G10,Childress_2013,Uddin_2017,Smith_2020,2021Kelsey}). This is applied as a step correction around $10^{10}M_{\odot}$, and is known as the ``{mass step}''. SNe Ia in low-mass galaxies are fainter after standard light curve correction compared to SNe Ia in high-mass galaxies.

However, the origin of the mass step remains poorly understood. One possibility is that the mass step is an artefact, driven by the overly simplistic single-valued colour correction that is applied to SN Ia luminosities (\citealp{brout2021dust, meldorf2022, Popovic_2023}). There are at least two physically plausible mechanisms for the colour correction. First, more energetic SNe Ia will be brighter, hotter and therefore bluer. This is intrinsic to the SN. Second, dust will redden and dim SNe Ia. This is extrinsic to the SN. There is a no priori reason to expect that the relationship between colour and luminosity should be the same for both, yet this is what is assumed. The consequence of this assumption might be the mass step.

Many galaxy properties correlate with the properties of the dust in them. For example, more massive galaxies are dustier (\citealp{2021Triani}), as are galaxies with higher specific star formation rates (sSFR) \citep{2017Orellana}, which is the star formation rate (SFR) normalised by stellar mass. These correlations then naturally lead to trends between Hubble residuals and the properties of the hosts if the colour correction is unable to capture both reddening by dust and the intrinsic colour-luminosity at the same time. For example, \cite{Dixon_2022} find a correlation between Hubble residuals and the Balmer decrement, a measure of extinction by dust.

Previous work by \cite{Dixon_2022} derived host galaxy properties using stacked spectra instead of broad band photometry to explore what is physically driving the mass step. The most significant correlation uncovered was with the equivalent width (EW) of the [OII] $\lambda\lambda$ 3727, 29 doublet which is an indicator of the sSFR.

In this paper, we build upon this finding by directly measuring the [OII] EW for each SN Ia host galaxy. By analysing these SNe Ia, we are able to derive correlations between their Hubble residuals and [OII] EWs. We find that the [OII] EW trend is more significant than the commonly used mass step correction, highlighting its potential for improving the standardisation of SNe Ia. Next, we explore the impact of applying our [OII] EW correction to a sample of 20 SNe Ia hosted in galaxies ($z$ $<$ 0.012) with Cepheid-derived distances, which are used in constraining $M^{B}_{0}$. Finally, applying these results to the Pantheon+ SNe Ia Hubble flow sample, we update the values $M^{B}_{0}$ and $H_{0}$.

\section{Hubble Flow Galaxies} \label{OzDES/DES}
\subsection{DES and OzDES}

The Dark Energy Survey (DES) ran for six observing seasons from 2013 to 2019 and used
the 570 megapixel Dark Energy
Camera (DECam; \citealp{Flaugher2015}) situated on the 4-metre
Victor M.~Blanco Telescope \citep{Abbott2016}.
DES has observed hundreds of millions of galaxies and discovered thousands of supernovae \citep{hartley2020dark}.
The data are passed through the DES Image Processing Pipeline \citep{Morganson_2018}, and SN transients are identified with a difference imaging pipeline \citep{Kessler_2015}.

The Australian Dark Energy Survey (OzDES) was undertaken over the same observing period as DES, using the 3.9-m Anglo-Australian Telescope (AAT) at Siding Spring Observatory, along with the AAOmega spectrograph and the 2dF fibre positioner (\citealp{Yuan2015, Childress_2017}). The 2dF fibre positioner allocates fibres, and accommodates up to 8 guide stars and 392 science targets within a 2.1 degree field, which aligns with the DECam imager's field of view.
The primary aims of OzDES were measuring redshifts of SN hosts, confirming the spectral type of SNe, and monitoring AGNs over a wide redshift range \citep{hoorman2019}.
The wavelength coverage is between 3700 and 8800$\angstrom$, and the faintest objects have an apparent magnitude $r\approx$ 24 mag.
The second OzDES data release contains 375,000 spectra of 39,000
objects and is described in \cite{Lidman_2020}.


\subsection{Sample Selection}

We use the DES5YR photometrically identified SNe Ia sample described in \cite{2022Anais} and updated in \cite{vincenzi2024dark} for the final cosmology analysis \citep{descollaboration2024dark}. SNe Ia were classified using the SuperNNova classifier \citep{2020Anais}.

\begin{table}
\centering
\caption{From the DES5YR photometric sample containing 1499 SNe Ia, we apply specific cuts that result in a final sample of 707 OzDES host galaxies for our analysis.}
\begin{tabular}{lr}
\hline
\textbf{Selection Cut} & \textbf{\# SNe Ia} \\ 
\hline
DES5YR SNe Ia sample & 1499 \\ 
\hline
OzDES host spectra & 1248 \\ 
\hline
Redshift reliability $>95\%$ & 1191 \\
\hline
Not contaminated by SN Ia light & 744 \\
\hline
EW [OII] $<$ 0 & \textbf{707} \\
\hline
\end{tabular}
\label{Table:cuts}
\end{table}

Initially, we have 1499 SNe Ia, where the probability of being a SNe Ia is greater than 0.5 \citep{descollaboration2024dark}. We then apply specific selection cuts to our sample of hosts (see Table \ref{Table:cuts}). Each DES transient is associated with its host galaxy, which is identified using the smallest directional light radius from the deep image stacks discussed in \citet{Wiseman_2020}. Spectroscopic redshifts are obtained from the OzDES global redshift catalogue \citep{Lidman_2020}, and cuts are made regarding the redshift reliability for each galaxy. We only consider host redshifts with a quality flag of 3 or higher. This depends on identifying either a prominent single feature or multiple weaker features, resulting in a confidence level exceeding 95$\%$. We also remove galaxies that may have been affected by supernova light contamination. Specifically, we exclude hosts that were observed within two months before or five months after the peak luminosity of the SN Ia explosion. After this cut, there are 744 SNe Ia. We then make a final cut by omitting hosts with positive [OII] EWs (see Section \ref{SN Ia Host Galaxy Properties}), resulting in a final sample of 707 SN Ia host galaxies for our analysis.


\subsection{Deriving Hubble Residuals}

Each DES5YR SN Ia light curve is fit using the SALT3 light curve model (\citealp{Kenworthy_2021, Taylor_2023}), which is based on SALT2 \citep{G10}. We note that the differences between using SALT2 and SALT3 have been found to have a minimal impact on SN Ia cosmology results \citep{Taylor_2023}. The standardisation parameters consist of the colour ($c$), stretch ($x_1$), and peak brightness ($m_x$). We can then derive the observed distance modulus ($\mu_{\mathrm{obs}}$) for each SNe Ia using the modified Tripp equation \citep{1998A&A...331..815T}:
\begin{equation}
\mu_{\text{obs}, i} = m_{x, i} + \alpha x_{1, i} - \beta c_i + \delta_{\mathrm{host}, i}  - M - \delta_{\mathrm{bias}, i},
\end{equation}
where $M$ is the SN Ia absolute magnitude, with $x_{1} = 0$ and $c = 0$. $\delta_{\mathrm{host}}$ represents an additional correction for observed correlations between the SN Ia peak brightness and host galaxy properties. Commonly this is expressed in the form of a ``mass step'', where $\delta_{\mathrm{host}}$ = $\gamma G_{\mathrm{host}}$. 
\begin{equation}
    G_\mathrm{host} = \left\{ \begin{array}{cl}
+1/2 & \mathrm{for} \ M_{*} > M_\mathrm{step} \\
-1/2 & \mathrm{otherwise},
\end{array} \right.
\end{equation}
where $M_{*}$ is the galaxy stellar mass, and $\gamma$ is the size of the mass step. The division point ($M_\mathrm{step}$) is commonly taken as $10^{10}$ $\mathrm{M_{\bigodot}}$. 
The global fitting parameters for the DES5YR analysis are $\alpha = 0.161\pm0.001$, $\beta = 3.12\pm0.03$, and $\gamma=0.038\pm0.007$ \citep{descollaboration2024dark}. The terms indexed by `$i$' denote parameters specific to individual SNe Ia. $\delta_{\mathrm{bias}}$ accounts for biases arising from selection effects and light curve fitting, and is obtained from simulations using ``Beams with Bias Correction'' (BBC: \citealp{Kessler_2017}). The bias correction depends on redshift, stretch, colour and host galaxy mass (`BBC4D': \citealt{Popovic_2021}).
The light curve fitting process is discussed in greater detail in \cite{descollaboration2024dark}.

\begin{figure}
    \centering
    \includegraphics[scale = 0.45]{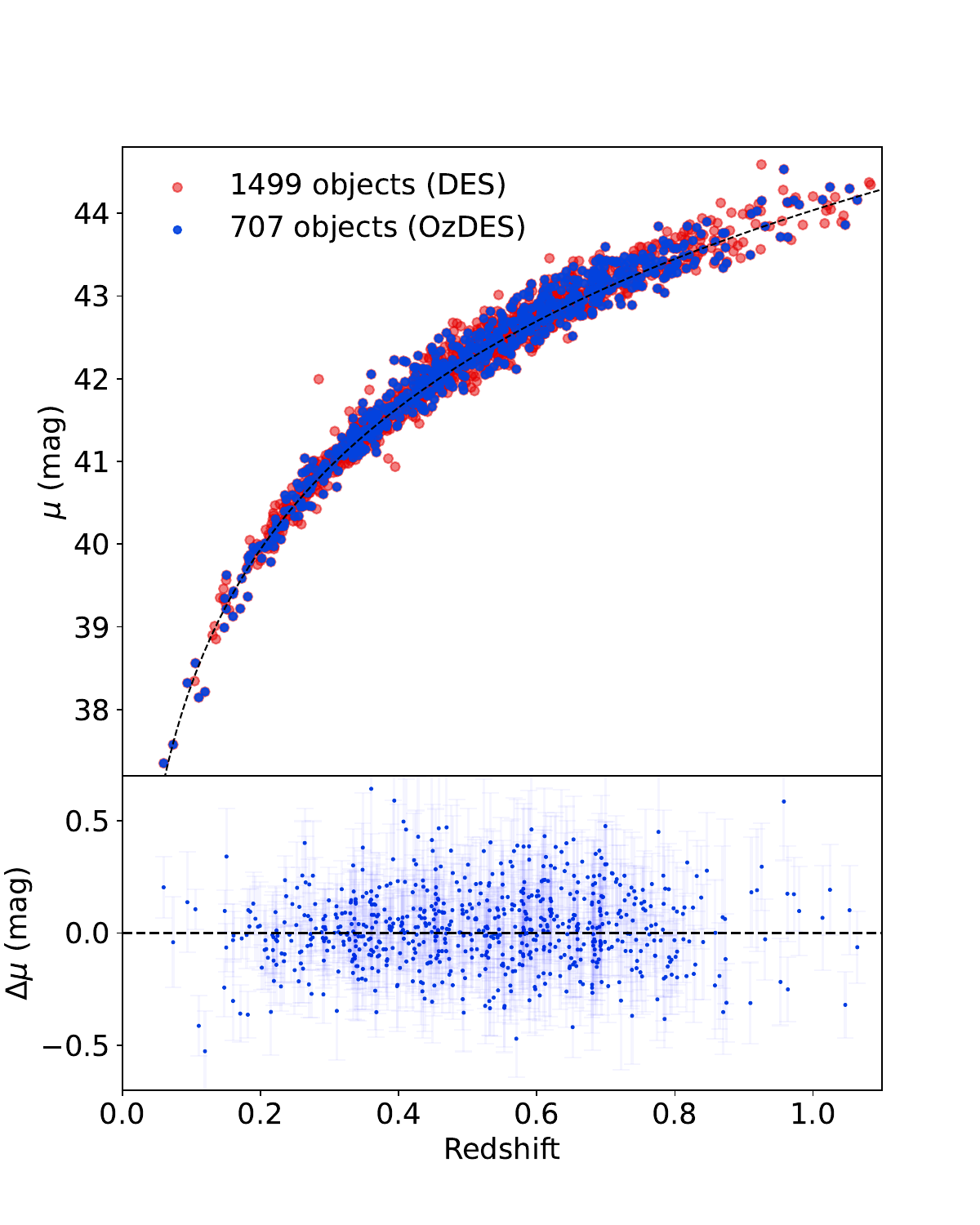}
    \caption{Upper plot: Hubble diagram showing the DES5YR sample containing 1499 SNe. After our selection cuts, we obtain a sample of 707 SNe Ia. Lower plot: Hubble residuals ($\Delta \mu$) for 707 SNe Ia.}
    \label{fig:DES_HD}
\end{figure}

The Hubble residuals ($\Delta \mu$) in our analysis represent the difference between the observed distance modulus ($\mu_{\text{obs}}$) and the distance modulus measured by DES using a flat $\Lambda$CDM cosmological model ($\mu_{\text{cosmo}}$), where $H_0 = 70\hubble$ and $\Omega_{\text{m}} = 0.352$ \citep{descollaboration2024dark}:
\begin{equation}
\Delta \mu = \mu_{\text{obs}} - \mu_{\text{cosmo}}.
\end{equation}
We then construct a Hubble diagram for our SNe Ia, as illustrated in Figure \ref{fig:DES_HD}. 

\begin{figure}
    \centering
    \includegraphics[scale=0.39]{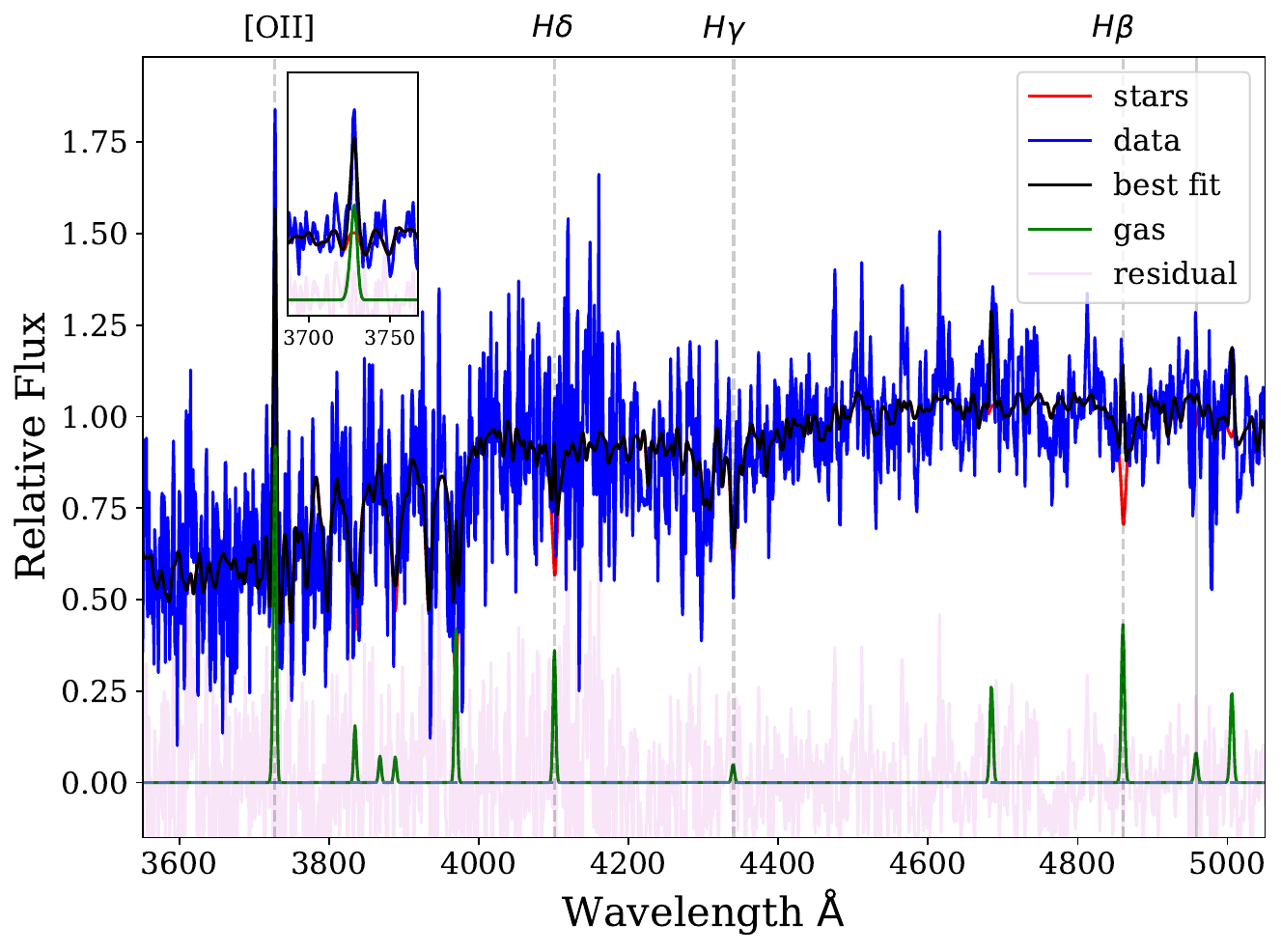}
    \caption{An example of using pPXF \citep{2017_Cappellari} to measure the [OII] EW for an OzDES galaxy. The blue line shows the galaxy spectrum, the black line is the best fit, the red line is the stellar continuum, the green line is the gas component, and the pink line illustrates the residual flux. We zoom in on the [OII] emission line to illustrate that this measurement ($-8.94\pm1.63$ $\angstrom$) is achievable even with a low signal-to-noise ratio (S/N) for the continuum.}
\label{ozdes_individual_galaxies}
\end{figure}

\subsection{SN Ia Host Galaxy Properties} \label{SN Ia Host Galaxy Properties}

We use penalised pixel-fitting (pPXF), which is a full spectrum fitting approach to extract stellar population parameters for each galaxy \citep{2004_Cappellari}. In building each synthetic spectrum, we utilise the E-MILES spectral library of single stellar populations \citep{Vazdekis_2016}, with stars older than 30 Myr, spanning a metallicity range of [M/H] from $-1.792$ to $+$0.26, and assuming a \cite{Salpeter_1955} initial mass function with a slope of 1.30 \citep{miles}. They were obtained at the 2.5-m Isaac Newton Telescope in Spain and cover the wavelength range 3525-7500\,$\angstrom$, 
with a spectral resolution of 2.5 $\angstrom$. To account for factors such as extinction and calibration errors, a fourth-order multiplicative polynomial was used to warp the spectral continuum. Full details of the pPXF methodology are given in \cite{2017_Cappellari}.

\begin{figure}
    \centering
    \includegraphics[scale = 0.50]{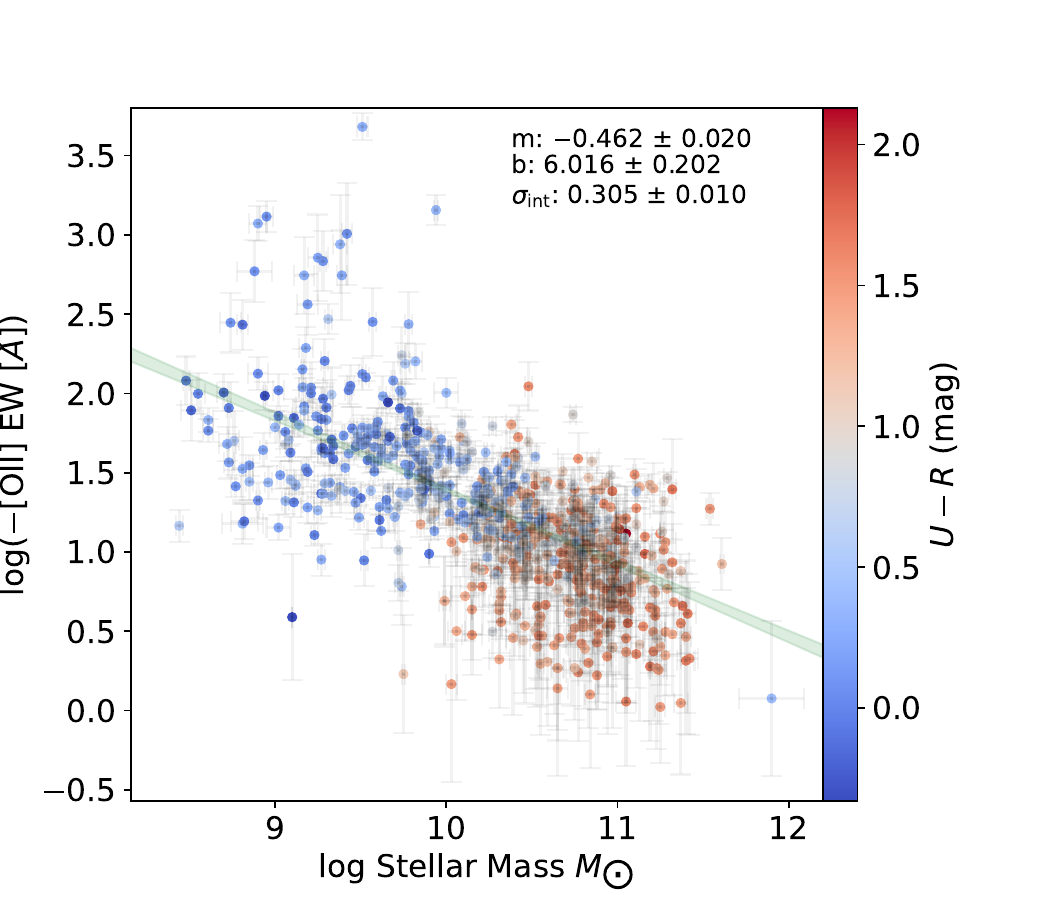}
    \caption{We find a strong correlation between host galaxy stellar mass and the log of the negative EW of [OII] for our sample of 707 galaxies. The slope of the linear fit is $-$0.462$\pm$0.020 and we find that more massive galaxies tend to have lower sSFRs.  Additionally, we add a comparison with $U-R$ rest-frame colour, showing more massive, passive galaxies tend to be redder.
    \label{fig:comparing_host_parameters}} 
    
\end{figure}

Building on the investigation in \cite{Dixon_2022}, we choose to focus on the [OII] emission line, given it has the strongest correlation obtained from the OzDES spectra and is an indicator of sSFR. Importantly, we now measure [OII] EW for each OzDES host galaxy. This is achieved by using pPXF to obtain the stellar continuum and emission line components. An example host galaxy spectrum, characteristically exhibiting a low signal-to-noise (S/N) of 3, is shown in Figure~\ref{ozdes_individual_galaxies}. Excluded from this analysis are objects with positive EWs. These are likely to be passive galaxies with little ongoing star formation.

We estimate uncertainties by perturbing each point in our spectra, where the magnitude of the perturbation is taken from a Gaussian that has a mean of zero and a variance that is determined from the variance spectrum. We run pPXF on 500 perturbed spectra and then measure the uncertainty in [OII] EW. 

Figure~\ref{fig:comparing_host_parameters} explores the relationship between host galaxy properties derived from photometry, which include stellar mass,  $U-R$ rest-frame colour (\citealp{descollaboration2024dark, vincenzi2024dark}) and our [OII] EW measurements. The plot highlights a strong trend between stellar mass and [OII] EW. Additionally, the comparison with $U-R$ shows that galaxies with higher stellar mass and a lower sSFR tend to be redder.

\begin{figure}
    \centering
    \includegraphics[scale=0.55]{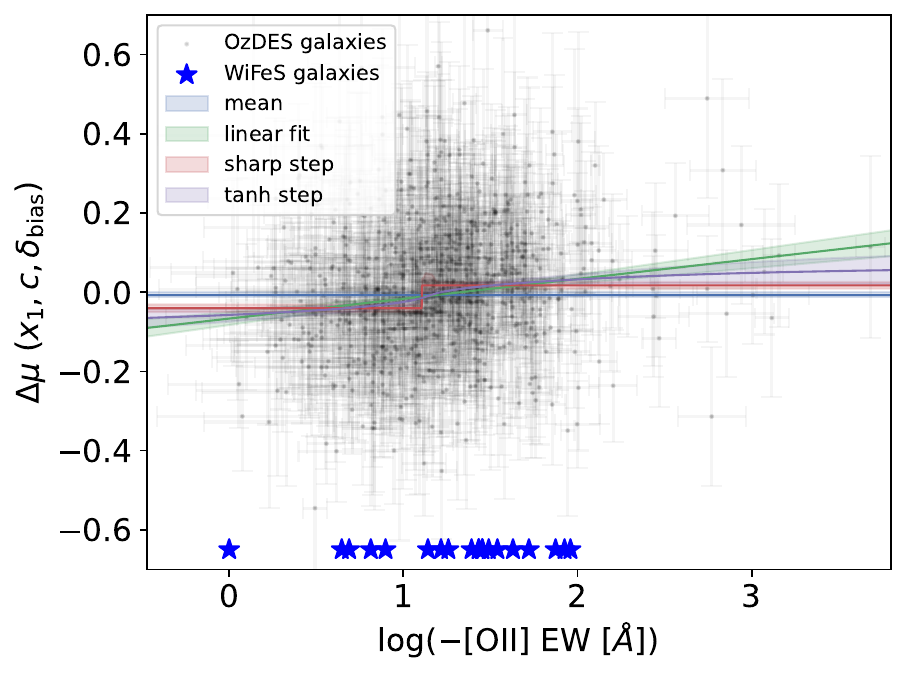}
    \caption{Host galaxy sample containing 707 objects. We plot [OII] EW against Hubble residuals ($\Delta \mu$), derived using stretch ($x_{1}$), colour ($c$) and bias corrections ($\delta_{\mathrm{bias}}$) for each SN Ia light curve. Compared to the final DES analysis \citep{descollaboration2024dark}, we omit the mass step correction ($\delta_{\mathrm{host}}$) to explore the potential of using an [OII] EW correction instead. A strong trend with a slope of 0.050$\pm$0.013 (4.0$\sigma$ significance) is evident, where fainter SNe Ia tend to be in hosts with larger [OII] EWs. Additionally, we examine the impact of different fitting functions (linear, step, smoothed step/tanh). The 20 SN Ia calibrator galaxies (blue stars) are plotted at a fixed $\Delta \mu =-0.65$ for reference and are discussed in Section~\ref{WiFeS}. We note that most of the SN Ia calibrators reside in galaxies with higher sSFRs.}
    \label{ozdes_fitting}
\end{figure}

\begin{figure*}
    \centering
    \includegraphics[width=0.8\textwidth]{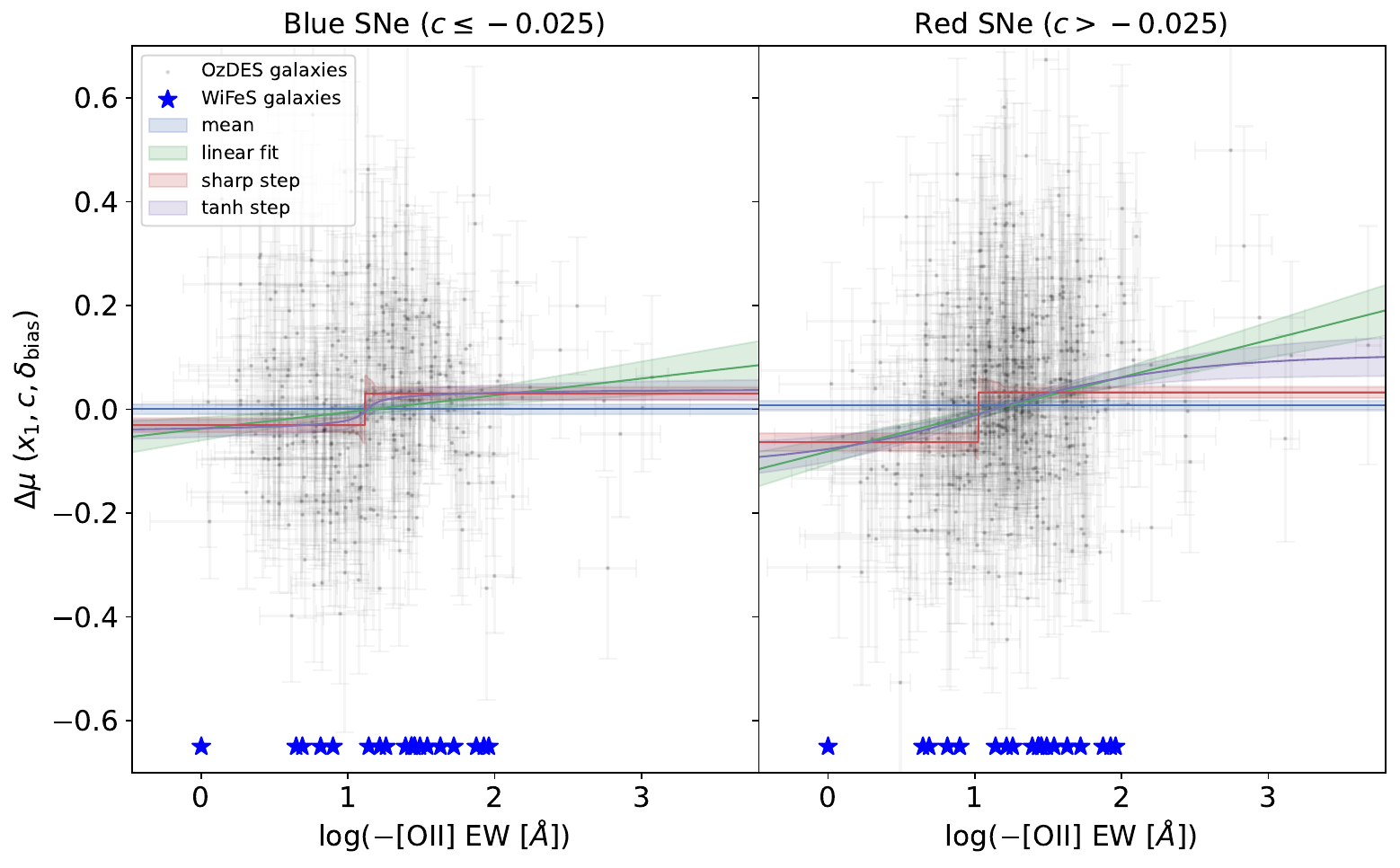}
    \caption{OzDES host galaxies split into sub-samples based on SN Ia colour, with 295 blue SNe Ia ($c < -0.025$) and 412 red SNe Ia ($c\geq -0.025$). We find that the trend seen in Figure \ref{ozdes_fitting} is more significant for redder SNe Ia (3.9$\sigma$), and is weaker for bluer SNe Ia (1.8$\sigma$).}
    \label{blue_red}
\end{figure*}


\subsection{Different Fitting Approaches}

Hubble residuals have been found to vary with host galaxy properties, and are typically characterised with a step function, such as the well-known ``mass step''. However, the physical origin of why such a step relation should exist is unclear. We can explore this systematic by using different fitting functions in obtaining our corrections. In this paper, we account for both $x$ and $y$ uncertainties in the fit. This allows us to determine the optimal fits for a linear trend, step function, and smoothed step function (hyperbolic tan).

In Figure \ref{ozdes_fitting}, we find a strong trend (4.0$\sigma$) between Hubble residuals, $\Delta {\mu}\,$($x_{1}$, $c$, $\delta_{\mathrm{bias}}$)\footnote{Hubble residuals that exclude the mass step correction but include stretch, colour, and bias corrections.}, and [OII] EW. Fainter objects after light curve and bias corrections tend to reside in galaxies that have higher [OII] EWs and therefore higher sSFRs. Additionally, we find that the smoothed step function prefers a significant amount of smoothing in obtaining the best fit. This converges towards the linear fit and away from the step function. Recent studies have found similar trends with Hubble residuals (\citealp{Rigault_2020, Briday_2022, Galbany_2022, Dixon_2022}). However, these studies employ different methods to derive the sSFR for each host galaxy (photometry vs. spectroscopy) and utilise different light curve correction parameters.


\subsection{Splitting by SN Ia Light Curve Properties}

We can also split our SNe Ia by colour, using the dividing line at $c = -0.025$ \citep{brout2021dust}. We find a noticeable difference between the red SNe ($c > -0.025$) and blue SNe $c\leq-0.025$), namely that redder SNe have a more significant trend with Hubble residuals and [OII] EW (3.9$\sigma$), compared to bluer SNe, where the correlation is weaker (1.8$\sigma$). \cite{2021Kelsey} found a similar result when splitting on $U-R$ rest frame colour. Overall these findings support the notion that galaxies hosting bluer SNe are less impacted by environmental dependencies and are better suited for use in cosmology \citep{Kelsey_2023}.


\section{SN Ia Calibrator Galaxies}\label{WiFeS}
\subsection{Observing and Data Reduction}
Our first step is to quantify host stellar population properties, with a focus on the [OII] EW measurements from nearby galaxies for which SNe Ia and Cepheid distances are available. Our sample is built from the recent analysis undertaken by the SH0ES collaboration (SNe, H0, for the Equation of State of Dark Energy) and is described in \cite{Riess_2022}. We note that the Cepheid calibration used by SH0ES relies on multiple distance anchors, including Gaia EDR3 parallaxes, masers in NGC 4258, and detached eclipsing binaries in the Large Magellanic Cloud. Additionally, the SH0ES galaxies also form part of the SN Ia sample used in the Pantheon+ analysis (\citealp{2022Brout, Scolnic_2022}), encompassing 1701 SN Ia light curves spanning the redshift range of $0.001 < z < 2.26$.

\begin{figure}
    \centering
    \includegraphics[scale=0.35]{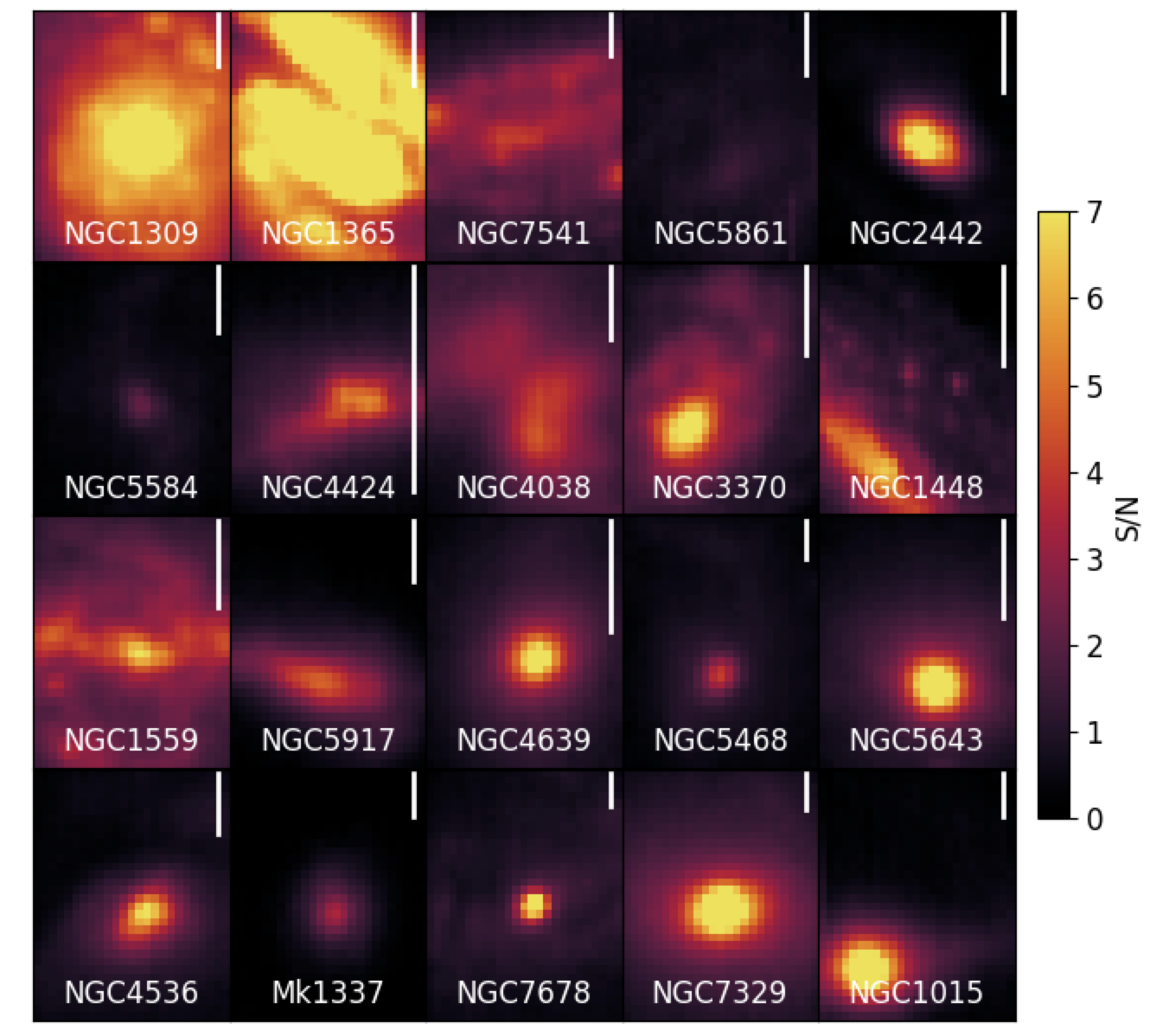}
    \caption{S/N maps for 20 SN Ia calibrator galaxies observed using WiFeS. The white line in each subplot represents a scale of $\sim$1 kpc, given each galaxy's redshift.}
    \label{fig:S/N}
\end{figure}

Initially, we selected a subset of 28 galaxies, limiting our selection to a declination $\leq$ 20 degrees for observing objects using the 2.3m ANU telescope at Siding Spring Observatory. We utilise the WiFeS instrument (\citealp{Dopita_2007, Dopita_2010}) which is a double-beam, image-slicing, integral-field spectrograph with a 25 by 38 arcsecond field of view (FoV). We use the RT560 dichroic beamsplitter along with the R3000/B3000 gratings, covering the wavelength range $3300-9200$ $\angstrom$. Our galaxies were observed in nod $\&$ shuffle mode, with two 1600s exposures, comprising of 800s on the object and 800s on sky positions respectively. The median seeing was 2 arcseconds. Due to time constraints, we observed 20 galaxies which are listed in Table \ref{WiFeS Galaxies Table}.

Our WiFeS data were reduced using the PyWiFeS data reduction pipeline \citep{Childress_2014}. The wavelength solution was derived using arc lamp exposures. Wire frames are obtained for spatial alignment of the slitlets while flat field exposures are used to remove pixel-to-pixel sensitivity variations. For each night, a standard star was acquired for flux calibration and to correct for telluric absorption. Typically, we observed a white dwarf spectrophotmetric standard. For each exposure, two data cubes from the blue/red CCD's are generated, containing spatial and spectral information across the FoV.

\subsection{Aperture Selection}
\begin{figure}
    \centering
    \includegraphics[scale=0.35]{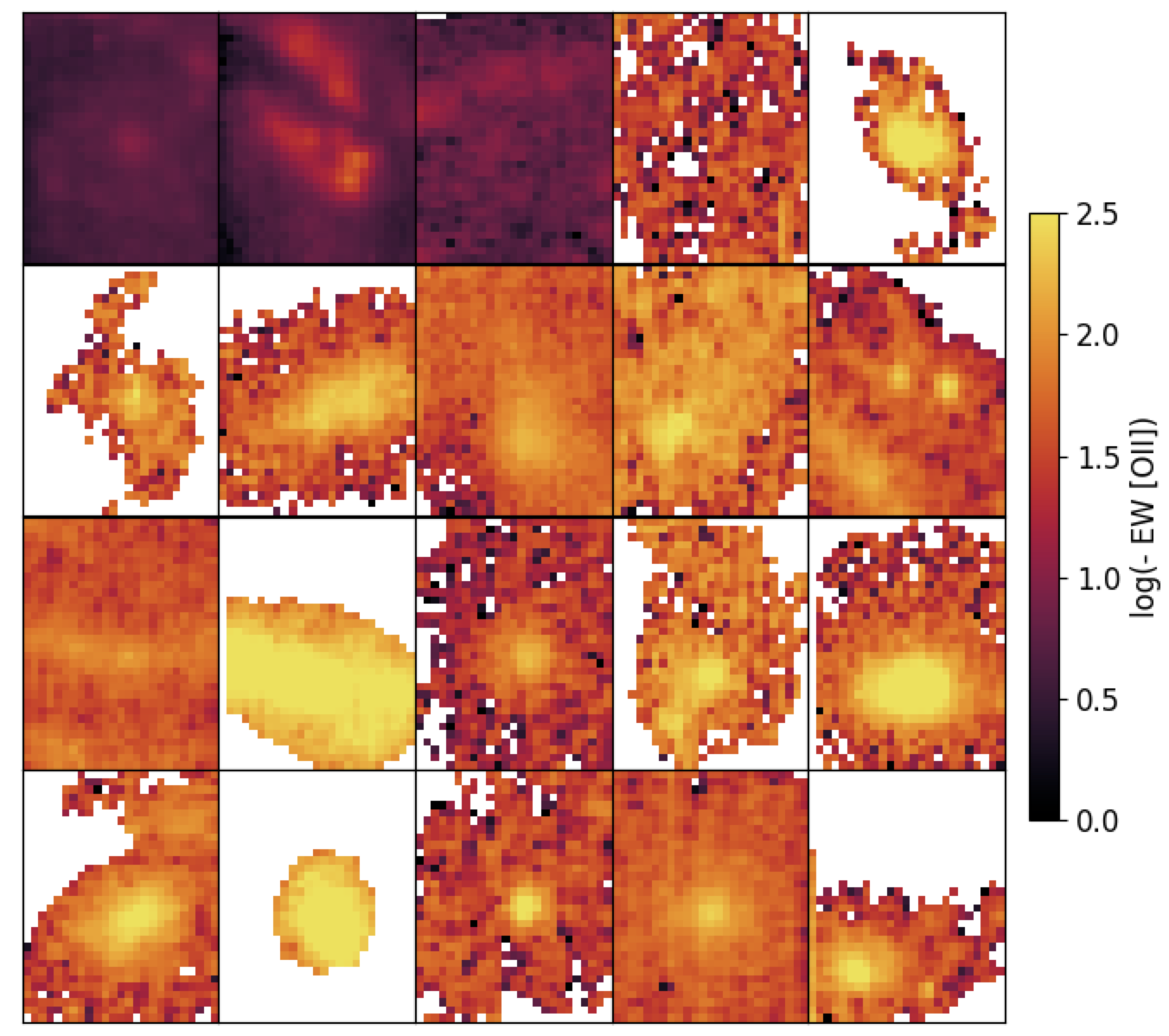}
    \caption{[OII] EW maps for each of the 20 SN Ia calibrator galaxies. A mask is applied to include
    spaxels with a signal-to-noise ratio (S/N) greater than 0.5. This additionally helps remove spaxels which are more heavily impacted by sky subtraction residuals.}
    \label{fig:EW OII}
\end{figure}

The next step is to extract a 1D spectrum from each WiFeS cube, with the goal of measuring integrated [OII] EW's that are comparable to those from OzDES. The OzDES sample spans the redshift range 0.12$-$1.06, with a median redshift of 0.52.  At these distances, the 2 arcsecond fibre aperture corresponds to a projected aperture $\approx$ 5$-$16 kpc (but with seeing effects on a similar scale). Recognising the risk of `aperture effects' creating a systematic difference between the low-$z$ and high-$z$ measurements, we use the largest aperture that the data can support, trying to make sure that we capture as much of the [OII] emission as possible, and being mindful of strong [OII] EW gradients (see Figures \ref{fig:S/N} $\&$ \ref{fig:EW OII}). A potential challenge associated with this selection relates to spaxel quality in regions near the aperture edges. To address this, we exclude poor-quality (S/N $< 0.5$) spaxels from our analysis. 

We can further explore the systematic uncertainty associated with our aperture selection through three different cases. First, we can select all of the available spaxels to best compare with the OzDES hosts. Second, we can make a cut based on the S/N of each spaxel. This effect introduces a bias towards spaxels with stronger lines. However, this bias is mitigated when the spaxels are combined. Third, we then explore the impact of selecting a smaller aperture, and combine the variation in each approach to obtain the systematic involving our selection choice. Overall, for the linear correction, the systematic impact remains consistent regardless of our aperture selection and only differs by 0.005 mag. The variation in step correction is more significant, which is expected for galaxies that fall around the step location. However, the impact is small ($\sim$ 0.01 mag). Aperture corrections in spectroscopic surveys can artificially increase the size of the mass step by 0.02 mag and its significance \citep{Galbany_2022}. Hence we adopt this as a systematic in our analysis.



Once extracted, the red and blue spectra are spliced together and we determine the redshift of the combined spectrum using MARZ \citep{Hinton_2016}. We note the spectra from the red arm are rebinned to match the spectral resolution of the blue arm. 


\begin{table*}
\centering
\caption{SN Ia calibrator galaxies observed using WiFeS. For each galaxy, we show the redshift, RA, DEC, SH0ES SN Ia absolute magnitude ($M^{B}_{0,i}$), [OII] EW, and host stellar mass. We note that for NGC1448, we take the weighted average of the two SNe Ia hosted by that galaxy \citep{Riess_2022}.}
\begin{tabular}{c|c|c|c|c|c|c|c|c|c}
\hline
Galaxy & $z$ & RA & DEC & $M^{B}_{0,i}$ (mag) & $\sigma_{M^{B}_{0,i}}$ & [OII] EW ($\angstrom$) & $\sigma_{\mathrm{OII}}$ & log Stellar Mass $\mathrm{M_{\bigodot}}$\\
\hline
NGC1015 & 0.0088 & 02h38m11.565s & $-$01d19m07.020s & $-19.220$ & 0.120 & $-1.44$ & 0.95 & 9.906 \\
NGC1309 & 0.0071 & 03h22m06.556s & $-$15d23m59.794s & $-19.337$ & 0.102 & $-20.09$ & 0.09 & 9.890 \\
NGC1365 & 0.0055 & 03h33m36.458s & $-$36d08m26.370s & $-19.479$ & 0.108 & $-27.58$ & 2.55 & 10.732 \\
NGC1448 & 0.0039 & 03h44m31.915s & $-$44d38m41.380s & $-19.199$ & 0.116 & $-28.08$ & 1.74 & 11.280 \\
NGC1559 & 0.0043 & 04h17m35.750s & $-$62d47m01.225s & $-19.361$ & 0.106 & $-31.25$ & 0.53 & 10.375 \\
NGC2442 & 0.0049 & 07h36m23.842s & $-$69d31m50.960s & $-19.223$ & 0.105 & $-23.74$ & 3.16 & 12.198 \\
NGC3370 & 0.0043 & 10h47m04.039s & $+$17d16m25.310s & $-19.186$ & 0.097 & $-6.80$ & 0.66 & 10.196 \\
NGC4038 & 0.0054 & 12h01m31.770s & $-$18d50m41.300s & $-19.207$ & 0.158 & $-44.29$ & 0.90 & 10.682 \\
NGC4424 & 0.0015 & 12h27m11.575s & $+$09d25m14.312s & $-19.369$ & 0.232 & $-21.89$ & 0.10 & 9.633 \\
NGC4536 & 0.0060 & 12h34m28.129s & $+$02d11m16.37s & $-19.287$ & 0.142 & $-50.76$ & 2.07 & 9.686 \\
NGC4639 & 0.0035 & 12h42m52.378s & $+$13d15m26.713s & $-19.364$ & 0.150 & $-6.95$ & 2.57 & 9.802 \\
Mk1337  & 0.0085 & 12h52m34.701s & $-$09d46m35.724s & $-19.267$ & 0.163 & $-91.50$ & 3.90 & 9.554 \\
NGC5468 & 0.0095 & 14h06m34.891s & $-$05d27m10.719s & $-19.127$ & 0.104 & $-42.62$ & 4.49 & 10.441 \\
NGC5584 & 0.0055 & 14h22m23.811s & $-$00d23m14.820s & $-18.971$ & 0.095 & $-19.49$ & 1.96 & 10.331 \\
NGC5643 & 0.0040 & 14h32m40.778s & $-$44d10m28.600s & $-19.362$ & 0.089 & $-64.53$ & 1.24 & 10.530 \\
NGC5861 & 0.0062 & 15h09m16.091s & $-$11d19m17.980s & $-19.287$ & 0.147 & $-11.84$ & 2.88 & 10.591 \\
NGC5917 & 0.0063 & 15h21m32.550s & $-$07d22m37.523s & $-19.284$ & 0.154 & $-157.51$ & 2.19 & 9.184 \\
NGC7329 & 0.0109 & 22h40m24.199s & $-$66d28m44.580s & $-19.244$ & 0.140 & $-2.20$ & 0.19 & 10.501 \\
NGC7541 & 0.0090 & 23h14m43.857s & $+$04d32m02.040s & $-19.086$ & 0.176 & $-30.28$ & 0.84 & 10.935 \\
NGC7678 & 0.0116 & 23h28m27.860s & $+$22d25m16.573s & $-19.106$ & 0.181 & $-11.91$ & 1.71 & 10.530 \\

\hline
\end{tabular}
\label{WiFeS Galaxies Table}
\end{table*}

\begin{figure*}
    \centering
    \includegraphics[scale=0.19]{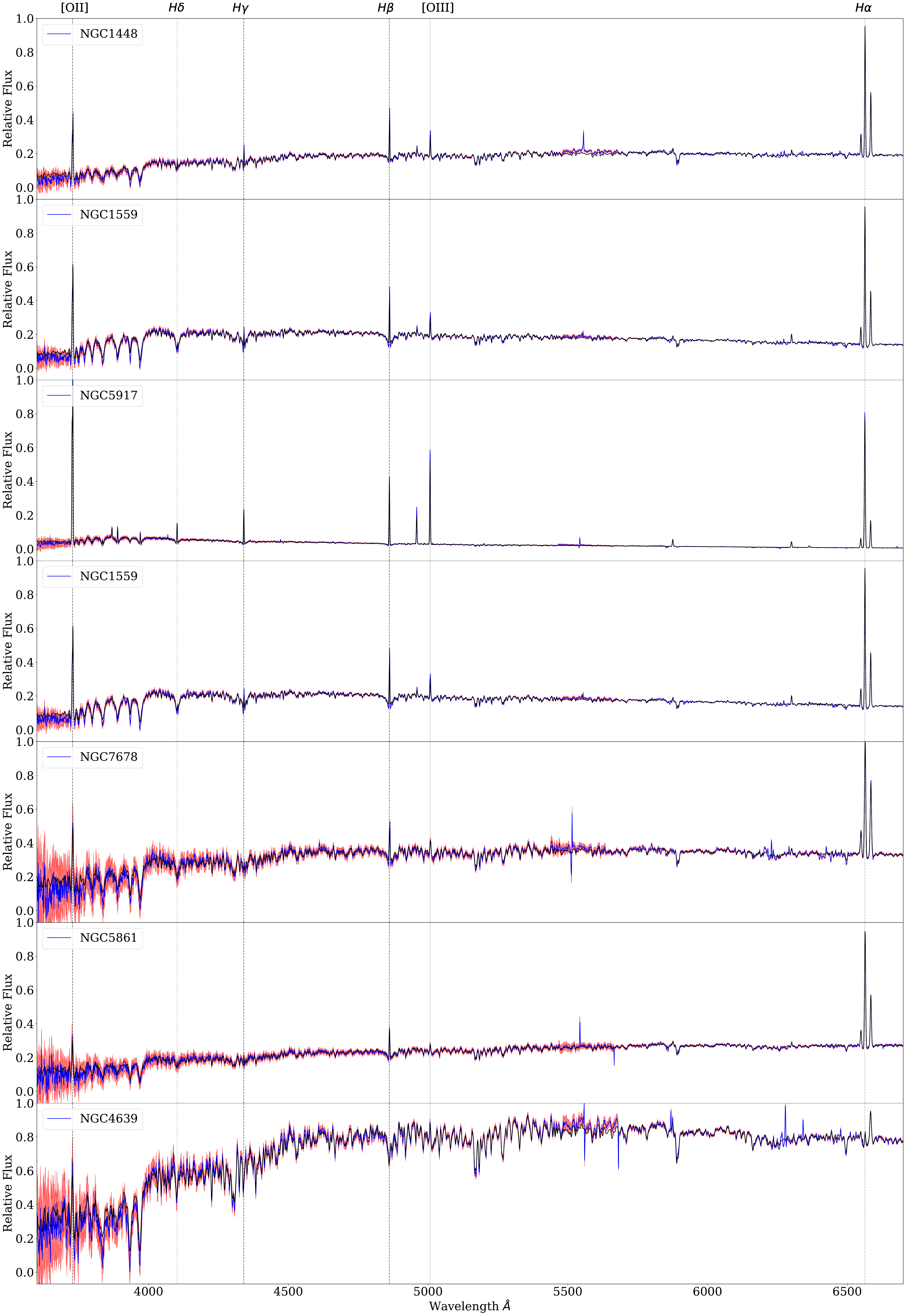}
    \caption{Examples of host galaxy spectra obtained using WiFeS. The original spectrum is in blue, the best fit is shown in black and is obtained using penalised-pixel fitting (pPXF; \citealt{Cappellari_2016}). The red region illustrates the standard deviation of the flux. Prominent spectral features that can be identified are [OII], [OIII], and the Balmer lines (H$\alpha$, H$\beta$, H$\gamma$, H$\delta$).}
    \label{fig:example_galaxies}
\end{figure*}

\section{Correcting SN Ia Absolute Magnitudes}\label{Correcting SN absolute brightness}

We now explore the impact of correcting SNe Ia for environmental dependence using [OII] EW. Similar to the OzDES hosts, we measure [OII] EW using pPXF for each of the 20 SN Ia calibrator galaxies observed using WiFeS (Figure \ref{fig:example_galaxies}). SN Ia absolute magnitudes ($M^{B}_{0,i}$) are obtained from the SH0ES analysis \citep{Riess_2022} and are listed in Table \ref{WiFeS Galaxies Table}. Hereafter, $\widetilde{M^B_{0,i}}$ represents the [OII] EW corrected absolute magnitude for each SN Ia, while $M^{B}_{0}$ denotes the weighted average of the 20 SN Ia calibrator galaxies. We note that our $M^{B}_{0}$ will differ from SH0ES, given that it is a subsample.

We determine our corrections by analysing the trends between $\Delta \mu$ and host galaxy properties (stellar mass, [OII], $U-R$). We explore these trends with a range of light curve correction parameters which include $x_{1}$, $c$, $\delta_{\mathrm{host}}$, $\delta_{\mathrm{bias}}$ and [OII]. 
\begin{itemize}
  \item Case 1 baseline: $\Delta {\mu}$ ($x_{1}$, $c$, $\delta_{\mathrm{bias}}$)
  \item Case 1: $\Delta {\mu}$ ($x_{1}$, $c$, [OII], $\delta_{\mathrm{bias}}$)
  \item Case 2 baseline: $\Delta {\mu}$ ($x_{1}$, $c$, $\delta_{\mathrm{host}}$, $\delta_{\mathrm{bias}}$)
  \item Case 2: $\Delta {\mu}$ ($x_{1}$, $c$, $\delta_{\mathrm{host}}$, [OII], $\delta_{\mathrm{bias}}$)
\end{itemize}
These corrections correspond to the rows in Figure \ref{fig:mass_oII_colour}, and summarised in Table \ref{Table:HR_corrections}. We discuss these results in more detail in Section \ref{Comparing_host_properties}.

Next we examine two specific cases that apply different [OII] EW corrections derived using the Case 1/2 baselines.

\subsection{Case 1: $\Delta {\mu}$ ($x_{1}$, $c$, [OII], $\delta_{\mathrm{bias}}$)} \label{Case 1}
\begin{figure*}
    \centering
    \includegraphics[width=0.90\textwidth]{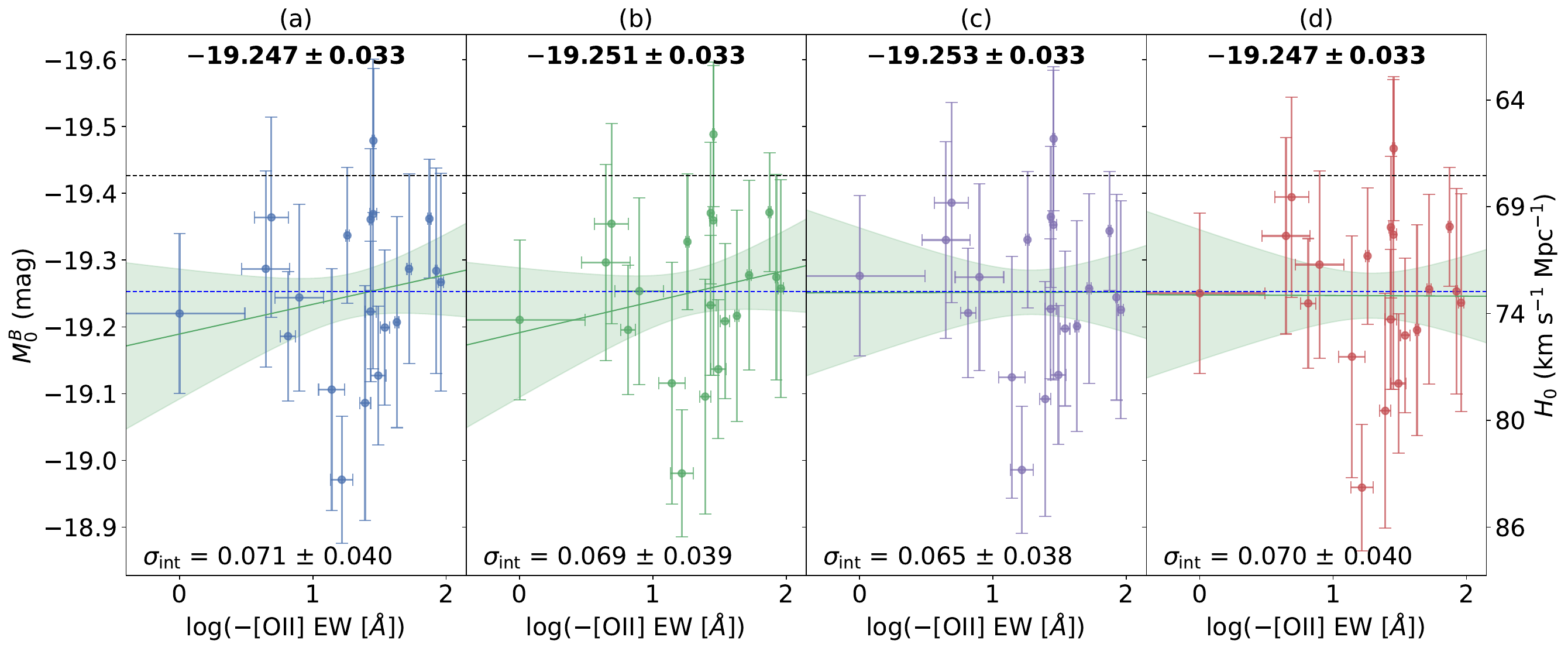}
\caption{Case 1: Exploring the impact of different [OII] EW corrections on SN Ia absolute magnitudes. (a) shows $M^{B}_{0,i}$ for each SN Ia calibrator galaxy with the Pantheon+ mass step applied, (b) is with the mass step removed, while (c) and (d) are with linear and step [OII] corrections respectively. We measure the intrinsic scatter ($\sigma_{\mathrm{int}}$) in each subplot and report the mean $M^{B}_{0}$, with values ranging from $-$19.253 to $-$19.247 mag. The [OII] correction helps minimise the trend between $M^{B}_{0}$ and [OII], as seen by comparing the linear fits (green). The dashed blue line represents the SH0ES analysis \citep{Riess_2022}, and the black dashed line corresponds to the Planck $H_{0}$ result \citep{Planck2018}.}
    \label{Correcting SN Ia hosts:Case1}
\end{figure*}

The first case involves replacing the mass step correction ($\delta_{\mathrm{host}}$) with our [OII] EW correction. We achieve this by removing $\delta_{\mathrm{host}}$ as described in \cite{descollaboration2024dark}, where $\gamma = 0.038$ mag, and are left with the Case 1 baseline which can now be applied as a linear correction using [OII] EW. However, we need to also remove the Pantheon+ mass step correction ($\delta^{P+}_{\mathrm{host}}$) for each of the SN Ia calibrators. We accomplish this by adding $\delta^{P+}_{\mathrm{host}}$ to $M^{B}_{0,i}$, where the size of $\gamma$ is set to 0.019 mag \citep{2022Brout} to align with the intrinsic scatter model (P21: \citealp{Popovic_2021}) used in the DES5YR analysis \citep{descollaboration2024dark}. We can then obtain $\widetilde{M^B_{0,i}}$ as is shown below:
\begin{equation}
    \widetilde{M^B_{0,i}} = M^{B}_{0,i} + \delta^{P+}_{\mathrm{host}} + (0.050\times\mathrm{[OII]}_{i}  - 0.066),
\end{equation}
where [OII] $= \log(-[\mathrm{OII}]$ EW).
Each step is shown in Figure \ref{Correcting SN Ia hosts:Case1}, where we first remove $\delta^{P+}_{host}$, then apply a linear and step [OII] EW correction, before we calculate $\widetilde{M^B_{0,i}}$ for each SN Ia calibrator galaxy.

After applying a [OII] EW linear correction, we measure $\widetilde{M^B_{0}}=-19.253\pm0.033$ mag. For a step correction, $\widetilde{M^B_{0}} = -19.247\pm0.033$ mag. We take the linear and step-fitting approaches to be the upper/lower extremes, defining the systematic range depending on our choice of fitting function. However, we do find that the linear fit is the preferred option, with the smallest deviance information criterion (DIC). Additionally, a smoothed step approach tends to converge to a linear trend, rather than a step.


\subsection{Case 2: $\Delta {\mu}$ ($x_{1}$, $c$, $\delta_{\mathrm{host}}$, [OII], $\delta_{\mathrm{bias}}$)} \label{Case 2}

\begin{figure*}
    \centering
    \includegraphics[width=1\textwidth]{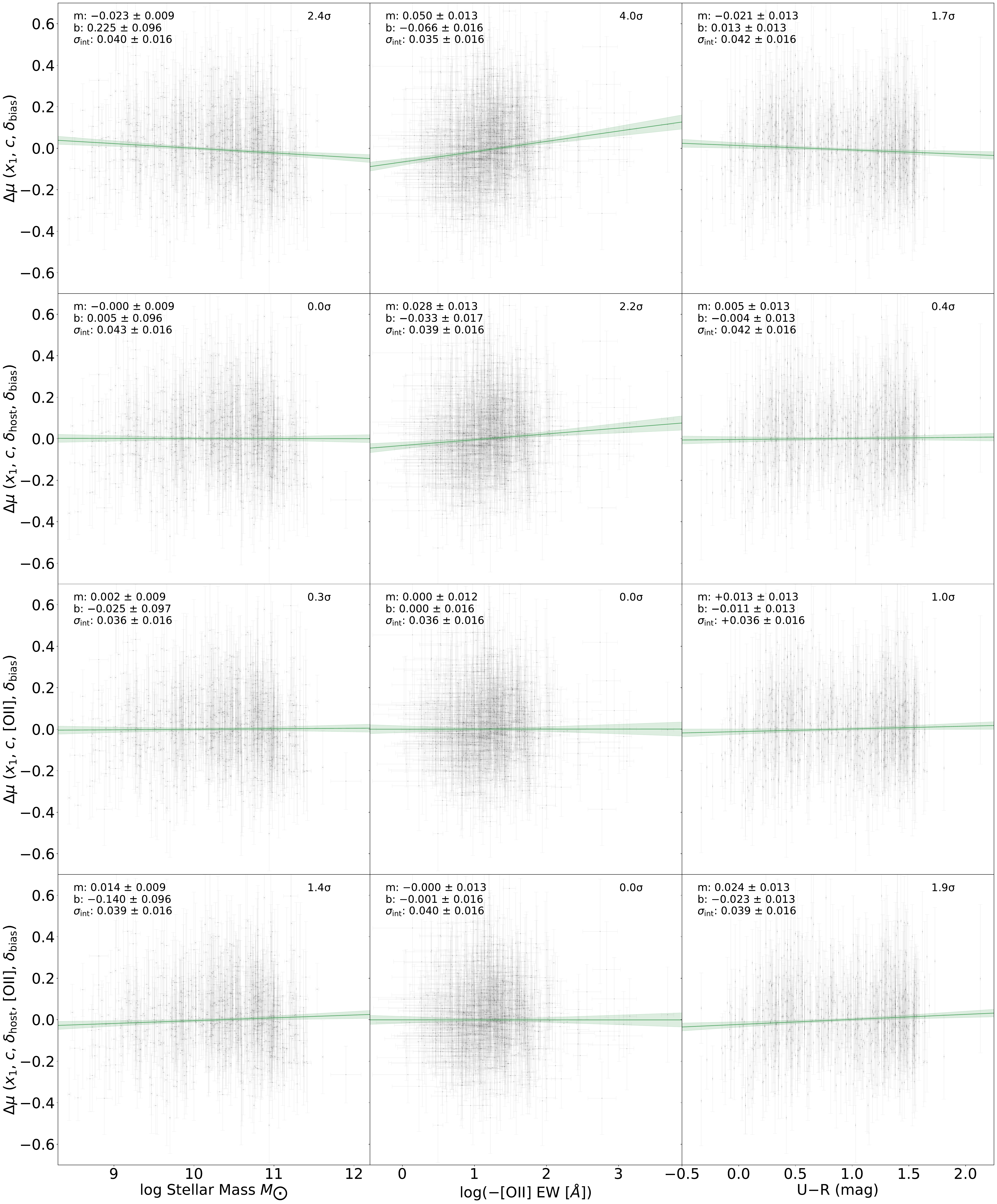}
    \caption{Comparing the effectiveness of host stellar mass, [OII] EW, and $U-R$ rest-frame colour as additional corrections. \textbf{First row:} $\Delta \mu$ are derived using stretch ($x_{1}$), colour ($c$) and bias corrections ($\delta_{\mathrm{bias}}$). We find that [OII] EW exhibits the most significant trend (4.0$\sigma$) and smallest intrinsic scatter. \textbf{Second row:} We then introduce a mass step correction $\delta_{\mathrm{host}}$ to obtain new $\Delta \mu$ values. Interestingly, a trend with [OII] EW persists (2.2$\sigma$), while stellar mass and $U-R$ colour show no significant trends. \textbf{Third row:} Here, we explore the case of replacing $\delta_{\mathrm{host}}$ with our [OII] EW correction. \textbf{Fourth row:} Lastly, we examine the case of utilising both a mass step and [OII] EW correction. We find that the best approach for minimising environmental dependence with $\Delta \mu$ across all three parameters ($\leq$ 1$\sigma$) involves a combination of $x_{1}$, $c$, [OII] and $\delta_{\mathrm{bias}}$.}
    \label{fig:mass_oII_colour}
\end{figure*}

In this case, we explore the impact of applying an additional [OII] EW correction to the mass-corrected DES5YR Hubble residuals. Even after applying $\delta_{\mathrm{host}}$ as a mass-step correction in the Case 2 baseline, our analysis in Figure \ref{fig:mass_oII_colour} (second row) reveals a 2.2$\sigma$ trend between the remaining $\Delta \mu$ and [OII] EW. We can then directly apply this correction to the SN Ia calibrators, and examine the impact on $M^{B}_{0,i}$:
\begin{equation}
    \widetilde{M^B_{0,i}} = M^{B}_{0,i} +(0.028\times\mathrm{[OII]}_{i}  - 0.033).
\end{equation}
Applying this [OII] EW correction results in a slight shift towards fainter $\widetilde{M^B_{0}}$ values which range from $-$19.244 to $-$19.239 mag. This approach, which combines the mass step and [OII] EW, proves to be less effective than Case 1 (fourth row, Figure \ref{fig:mass_oII_colour}).
We find that replacing the mass step directly with our [OII] EW correction (Case 1) is the better choice for mitigating trends between Hubble residuals and host galaxy properties (third row, Figure \ref{fig:mass_oII_colour}). Further discussion on the effectiveness of these host galaxy parameters is provided in Section \ref{Comparing_host_properties}.

\begin{table*} 
\centering 
\caption{Summary of the trends observed between $\Delta \mu$ and host galaxy properties for a range of light curve corrections (see Figure~\ref{fig:mass_oII_colour}). These parameters consist of $x_{1}$, $c$, $\delta_{\mathrm{host}}$, $\delta_{\mathrm{bias}}$ and [OII].}
\begin{tabular}{c|c|c|c}
\hline
$\Delta \mu$ & Slope & Significance ($\sigma$)  & $\sigma_{\mathrm{int}}$ \\
\hline
\multicolumn{4}{c}{\textbf{Stellar Mass}} \\
\hline
$x_{1}$, $c$, $\delta_{\mathrm{bias}}$ & $-0.023 \pm 0.009$ & $2.4$ & $0.040$ \\
$x_{1}$, $c$, $\delta_{\mathrm{host}}$, $\delta_{\mathrm{bias}}$ & $-0.000 \pm 0.009$ & $0.0$ & $0.043$ \\
$x_{1}$, $c$, [OII], $\delta_{\mathrm{bias}}$ & $0.002 \pm 0.009$ & $0.3$ & $0.036$ \\
$x_{1}$, $c$, $\delta_{\mathrm{host}}$, [OII], $\delta_{\mathrm{bias}}$ & $0.014 \pm 0.009$ & $1.4$ & $0.039$ \\
\hline
\multicolumn{4}{c}{\textbf{[OII]}} \\
\hline
$x_{1}$, $c$, $\delta_{\mathrm{bias}}$ & $0.050 \pm 0.013$ & $4.0$ & $0.035$ \\
$x_{1}$, $c$, $\delta_{\mathrm{host}}$,  $\delta_{\mathrm{bias}}$ & $0.028 \pm 0.013$ & $2.2$ & $0.039$ \\
$x_{1}$, $c$, [OII], $\delta_{\mathrm{bias}}$ & $0.000 \pm 0.012$ & $0.0$ & $0.036$ \\
$x_{1}$, $c$, $\delta_{\mathrm{host}}$, [OII], $\delta_{\mathrm{bias}}$ & $-0.000 \pm 0.013$ & $0.1$ & $0.040$ \\
\hline
\multicolumn{4}{c}{$\textbf{\textrm{U--R}}$} \\
\hline
$x_{1}$, $c$, $\delta_{\mathrm{bias}}$ & $-0.021 \pm 0.013$ & $1.7$ & $0.042$ \\
$x_{1}$, $c$, $\delta_{\mathrm{host}}$, $\delta_{\mathrm{bias}}$ & $0.005 \pm 0.013$ & $0.4$ & $0.042$ \\
$x_{1}$, $c$, [OII], $\delta_{\mathrm{bias}}$ & $0.013 \pm 0.013$ & $1.0$ & $0.036$ \\
$x_{1}$, $c$, $\delta_{\mathrm{host}}$, [OII], $\delta_{\mathrm{bias}}$ & $0.024 \pm 0.013$ & $1.9$ & $0.039$ \\
\hline
\end{tabular}
\label{Table:HR_corrections}
\end{table*}



\section{Discussion}
\subsection{Impact on $H_{0}$}

We can now use our corrected $M^{B}_{0}$ to make a revised estimate of $H_{0}$. We define the luminosity distance ($d_{L}$), which is commonly expressed using the distance modulus ($\mu$), the apparent ($m$) and absolute magnitudes ($M_{0}$) as shown below.
\begin{equation}
    \mu = m - M_0 = 5 \log_{10}(d_L) + 25.
\end{equation}
We then use a kinematic expression of our cosmological model which gives the $d_{L}$ as a function of redshift. The expression below is derived from a Taylor expansion of the Hubble–Lema\^{i}tre law for a flat universe, consisting of additional parameters, the cosmic deceleration, $q_0 = -\ddot{a}{\dot{a}^{-2}}a$ and jerk, $j_0 = -\dddot{a}{\dot{a}^{-3}}{a^2}$, where a is the scale factor and the dots are derivatives with respect to cosmic time. Given our current cosmological model, $q_0 = -0.55$ and $j_0 = 1.0$. We can neglect the higher-order terms and obtain the expression:
\begin{equation} \label{luminosity distance}
    \begin{aligned}
        d_{L} &\approx \frac{cz}{H_{0}} \left( 1 + \frac{1}{2}(1 - q_{0})z + \frac{1}{6}(1 - q_0 - 3q_{0}^2 + j_{0})z^2 \right) \\
        &= \frac{cz}{H_{0}}Q(z).
    \end{aligned}
\end{equation}

\begin{table*}
\centering
    \caption{Impact of different host galaxy corrections on the SN Ia absolute magnitude and $H_{0}$. We compare the two cases discussed in Section \ref{Correcting SN absolute brightness}. With no additional correction for our galaxies, we measure $M^{B}_{0} =  -19.247\pm0.033$ mag and $H_{0} = 73.24\pm1.11\hubble$. We also show the slope and significance of the trends in each case before and after corrections are implemented.}
\begin{tabular}{c|c|c|c|c|c}
\hline
 OzDES Hubble Flow SNe Ia & WiFeS Calibrator SNe Ia & Slope & Sig ($\sigma$) & $M^{B}_{0}$ (mag) & $H_{0}$ (\hubble) \\
\hline
& $M^{B}_{0}$ (SH0ES) & & & $-$19.253$\pm0.027$ & 73.04$\pm$1.01 \\
& $M^{B}_{0}$ & $-$0.044$\pm$0.071 & 0.62 & $-$19.247$\pm0.033$ & 73.24$\pm1.11$ \\
\hline
Case 1: $\Delta {\mu}$ ($x_{1}$, $c$, [OII], $\delta_{\mathrm{bias}}$) & $M^{B}_{0}$ $+$ $\delta^{P+}_{\mathrm{host}}$ & $-$0.048$\pm$0.072 & 0.67 & $-$19.251$\pm0.033$ & 73.11$\pm1.11$ \\
& $M^{B}_{0}$ $+$ $\delta^{P+}_{\mathrm{host}}$ $+$ $\mathrm{[OII]_{linear}}$ & $-$0.003$\pm$0.070 & 0.05 & $-$19.253$\pm0.033$ & 73.04$\pm1.11$ \\
& $M^{B}_{0}$ $+$ $\delta^{P+}_{\mathrm{host}}$ $+$ $\mathrm{[OII]_{step}}$ & 0.001$\pm$0.072 & 0.01 & $-$19.247$\pm0.033$ & 73.24$\pm1.11$ \\
\hline
Case 2: $\Delta {\mu}$ ($x_{1}$, $c$, $\delta_{\mathrm{host}}$, [OII], $\delta_{\mathrm{bias}}$) & $M^{B}_{0}$ $+$ $\mathrm{[OII]_{linear}}$ & $-$0.018$\pm$0.072 & 0.25 & $-$19.244$\pm0.034$ & 73.34$\pm1.15$ \\
& $M^{B}_{0}$ $+$ $\mathrm{[OII]_{step}}$ & $-$0.011$\pm$0.073 & 0.14 & $-$19.239$\pm0.033$ & 73.51$\pm1.11$ \\
\hline
\end{tabular}
\label{Table:hubble_constant}
\end{table*}

Substituting into equation \ref{luminosity distance}, we obtain a simplified expression to estimate $H_{0}$:
\begin{equation}
    \mathrm{log}_{10}H_{0} = \frac{M^B_{0} + 5a_B + 25}{5},
\end{equation}
\begin{equation}
    a_B = \log_{10}cz + \log_{10}Q(z) - \frac{m}{5},
\end{equation}
where $M^{B}_{0}$ is the absolute magnitude of the SN Ia calibrators, 
$a_B$ is the intercept of the distance-redshift relation, and we take $a_B = 0.714158$ which is the baseline value used in the SH0ES analysis \citep{Riess_2022}.


Without applying any additional correction, we measure $M^{B}_{0} =  -19.247\pm0.033$ mag and $H_{0} = 73.24\pm1.11 \hubble$. Importantly, depending on the chosen fitting approach and application of our [OII] EW correction (detailed in Cases 1 $\&$ 2), we find a difference of 0.47$\hubble$ with $H_{0}$ values ranging from 73.04 to 73.51$\hubble$. These results are summarised in Table \ref{Table:hubble_constant}.

As discussed in Section \ref{Case 2}, we find that the more effective approach in reducing the environmental dependence with Hubble residual is to directly replace the mass step correction with a linear [OII] EW correction. Subsequently, we adopt $M_{0} = -19.253\pm0.033$ mag, and determine $H_{0} = 73.04\pm1.11$\hubble as our nominal result. Ultimately, the shift in $H_{0}$ (up by 0.37$\%$ or down by 0.27$\%$) is small regardless of the applied [OII] EW correction. This suggests that environmental dependence on SN Ia brightness has a limited effect on reconciling the discrepancy with the CMB result for $H_{0}$, and a significant tension ($\sim$5$\sigma$) between these measurements persists.

\begin{figure*}
    \centering
    \includegraphics[width=0.9\textwidth]{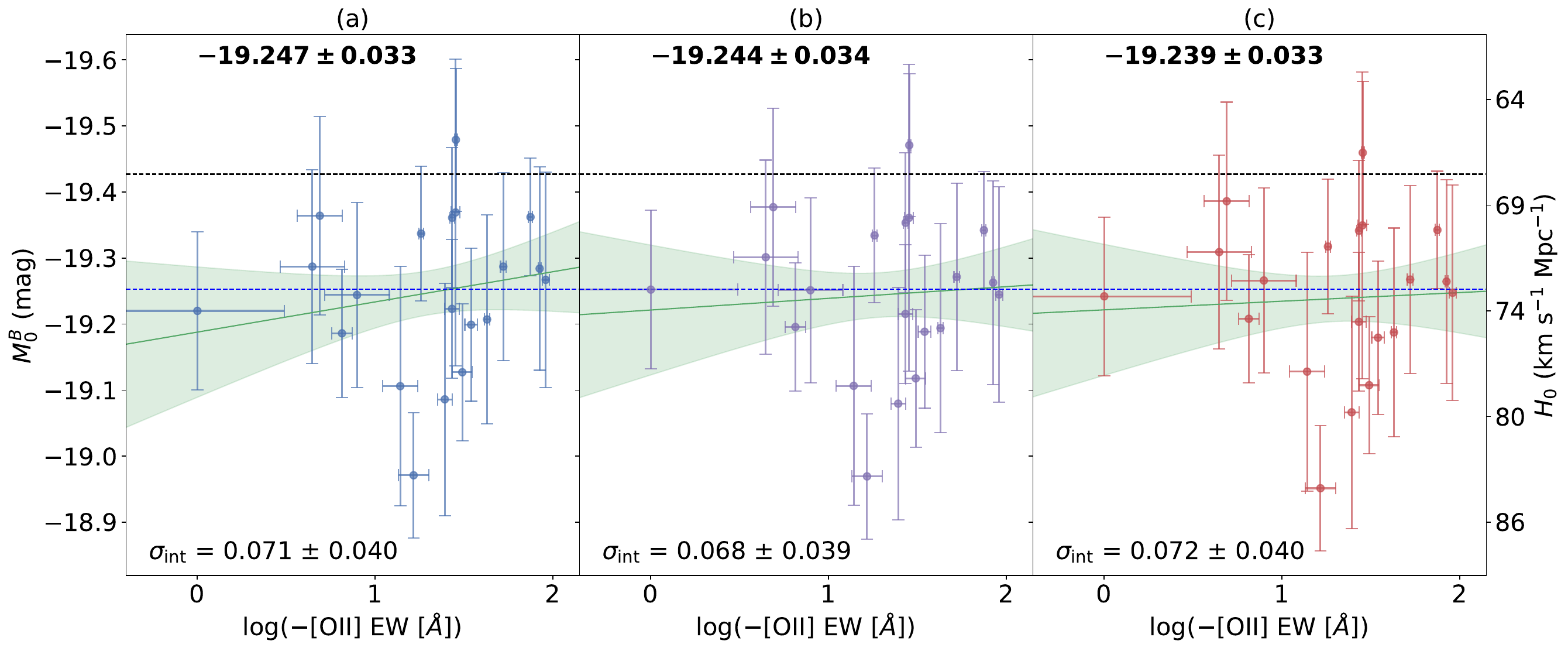}
    \caption{Case 2: Similar to Figure \ref{Correcting SN Ia hosts:Case1}, but here we only apply [OII] corrections to $M^{B}_{0,i}$, where the mass step has been corrected for. We find mean $M^{B}_{0}$ values ranging between $-$19.247 and $-$19.239 mag. While the [OII] correction reduces the trend between $M^{B}_{0}$ and [OII], its effectiveness is less than in Figure \ref{Correcting SN Ia hosts:Case1}.}
    \label{Correcting SN Ia hosts:Case2}
\end{figure*}

\begin{table} 
\centering 
\caption{Summary of the systematic error contributions for Case 1 (see Section \ref{Case 1}). This includes the selection of fitting functions used to derive [OII] EW correction, uncertainties in our choice of aperture, and the impact of not applying the [OII] EW correction to SNe Ia used for the third rung of the distance ladder. We estimate the total systematic uncertainty by taking the quadrature sum of these contributions.}
\begin{tabular}{c|c|c}
\hline
Case 1 & $\Delta$$M^{B}_{0}$ (mag)  & $\Delta$$H_{0}$ (km\,s$^{-1}$Mpc$^{-1}$) \\ 
\hline
Fitting Approach & 0.007 & 0.23
\\
Aperture Selection & 0.020 & 0.66 \\
Differential Impact & 0.009 & 0.30
\\
\hline
Total & 0.023 & 0.76\\
\hline
\end{tabular}
\label{Table:systematics_budget_case1}
\end{table}

\begin{table} 
\centering 
\caption{Similar to Table \ref{Table:systematics_budget_case1} but here we show the systematic budget for Case 2 (see Section \ref{Case 2}). We note the size of the [OII] EW correlation in Case 2 compared to Case 1, results in a reduced differential impact of 0.005 mag and hence a smaller overall systematic uncertainty.}
\begin{tabular}{c|c|c}
\hline
Case 2 & $\Delta$$M^{B}_{0}$ (mag)  & $\Delta$$H_{0}$ (km\,s$^{-1}$Mpc$^{-1}$) \\ 
\hline
Fitting Approach & 0.004 & 0.14
\\
Aperture Selection & 0.020 & 0.66 \\
Differential Impact & 0.005 & 0.17
\\
\hline
Total & 0.021 & 0.70\\
\hline
\end{tabular}
\label{Table:systematics_budget_case2}
\end{table}

\subsection{Systematic uncertainties}
An important bias affecting $H_{0}$ measurements, as identified by \cite{Rigault_2015}, arises from systematic differences in distances derived for SNe Ia in passive and star-forming environments. In our analysis, this bias arises after applying an [OII] EW correction derived from the OzDES hosts to SNe Ia in the calibrator sample (second rung) and not to the SNe Ia in the Hubble flow (third rung). The Hubble flow Pantheon+ SNe Ia, which has been corrected for the mass step, lacks spectra of all the host galaxies. Although we cannot apply our [OII] EW correction due to the lack of [OII] EW measurements, we can utilise the OzDES hosts ($z < 0.4$) to assess the potential impact of this bias.

To explore the impact of our [OII] EW correction across the second and third rungs, we can determine the differential impact on the variation in SNe Ia luminosities. The bias can be estimated by multiplying the relative differential fraction of host galaxy properties between the second and third rungs by the slope of the correlation with Hubble residuals, eg. $\mathrm{[OII]}_{\mathrm{slope}}\times(\mathrm{[OII]^{mean}_{second}}-\mathrm{[OII]^{mean}_{third}})$. The mean log($-$[OII] EW) of the WiFeS calibrator galaxies is 1.30, while that of the OzDES hosts ($z<0.4$) is 1.13. Using the slope measured in Figure \ref{ozdes_fitting}, the impact of this bias is $\sim 0.009$ mag in $M^B_{0}$.

To minimize the differential impact and achieve greater consistency with SH0ES, we select the top 80$\%$ of star-forming OzDES hosts. This selection mirrors the focus on star-forming hosts in the SH0ES analysis, ensuring a more comparable sample. The new OzDES sample has a mean log($-$[OII] EW) of 1.36, closer to that of the calibrator galaxies, and gives a reduced impact  $\sim 0.003$ mag in $M^B_{0}$.
While the smaller size of the applied correction mitigates the differential impact of different stellar populations on $H_{0}$ measurements, selecting SN Ia hosts with similar properties across the different rungs can further reduce this impact.

In Tables \ref{Table:systematics_budget_case1} $\&$ \ref{Table:systematics_budget_case2} we show the systematic errors in our analysis. One of the main differences between Case 1 and Case 2, is the differential impact between the different rungs, which is driven by the size of the applied correction.  Ultimately, we find that the impact of this systematic in our analysis is minimal.


\subsection{Which is more important, host galaxy mass or [OII] EW?} \label{Comparing_host_properties}
Galaxies are complex systems with many factors driving trends and potentially being intertwined. Ultimately, we want to explore whether stellar mass or [OII] is dominating the correlation with Hubble residuals, or if the two factors are strongly covariant.

In Figure~\ref{fig:mass_oII_colour}, we compare the effectiveness of host stellar mass and [OII] EW as additional SN Ia light curve parameters. This analysis is performed by applying different light-curve correction parameters specific to each galaxy property (Table~\ref{Table:HR_corrections}). In the first row, we remove the mass step correction ($\delta_{\mathrm{host}}$) as defined in \cite{descollaboration2024dark}. We find that [OII] EW exhibits the most significant trend (4.0$\sigma$) compared with stellar mass and $U-R$ colour. In the second row, a mass step has been applied and along with the `BBC4D' bias correction \cite{Popovic_2021}.

We find that the slope is flat for both stellar mass and $U-R$. However, we uncover a 2.2$\sigma$ trend with [OII] EW, which suggests that the correlation has not been entirely removed. In the third row, we instead directly apply our [OII] EW correction as an alternative to implementing a mass step correction. We find this correction is more effective than a mass step, reducing the significance of the trends across all three galaxy parameters to $\leq$ $1\sigma$. Additionally, we find that this approach results in the smallest intrinsic scatter, further suggesting that [OII] EW does better than the mass step correction. In the fourth row, we also examine the case with all light curve fitting parameters, but find that this approach is not as effective.

\begin{figure*}
    \centering
    \includegraphics[width=\textwidth]{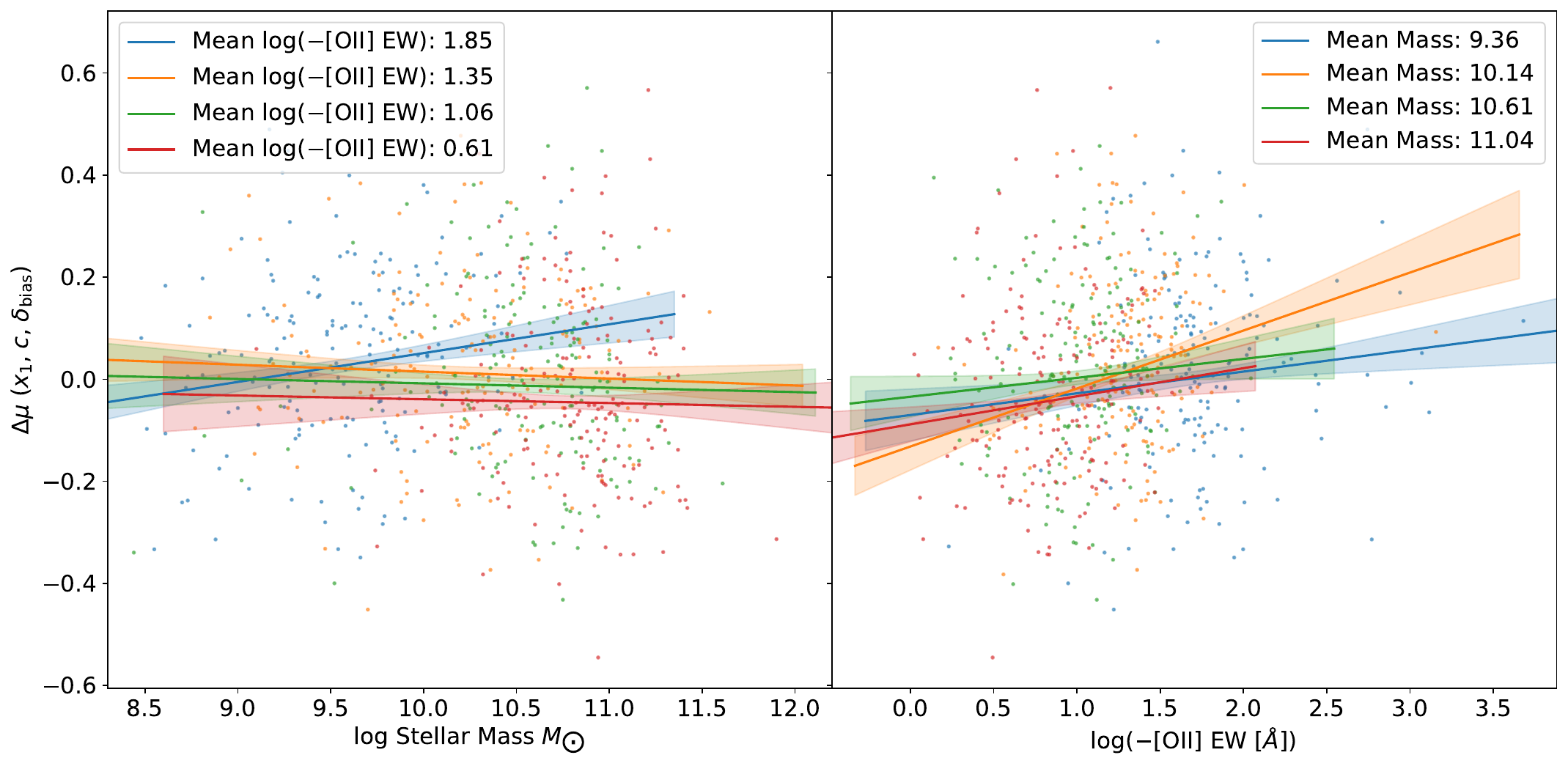}
    \caption{Left: First, we split the OzDES host galaxies into four [OII] EW bins with roughly the same number of objects. We then plot $\Delta \mu$ against stellar mass for each of the bins.
    Right: Similar, but instead we split by stellar mass, and then plot $\Delta \mu$ against [OII] EW.}
    \label{fig:mass vs oII}
\end{figure*}

To further probe these trends, we then split our sample by stellar mass or [OII] EW, with roughly the same number of objects in each bin. Then for each of the bins, we plot Hubble residual against the other galaxy parameter ([OII] EW/stellar mass). In Figure \ref{fig:mass vs oII}, we find that when splitting our sample in bins of [OII] EW, the lines differ. This suggests that the relationship between Hubble residual and stellar mass varies for galaxies with different sSFRs. When instead binning by stellar mass, the lines are more similar and overlap to a greater extend. This suggests that [OII] EW, as an indicator of sSFR, is more prominent in driving the correlation with Hubble residual for our OzDES host galaxies.

\subsection{$U-R$ and [OII] EW}
\begin{figure}
    \centering
    \includegraphics[scale = 0.5]{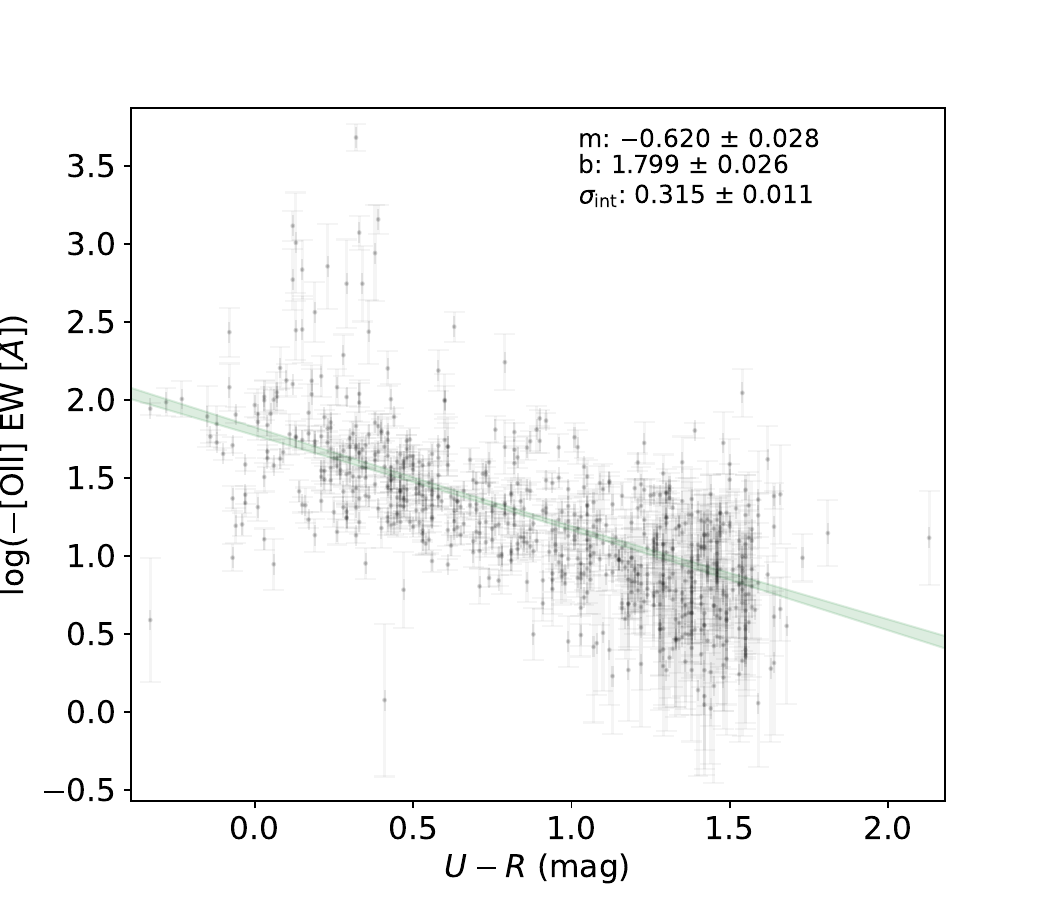}
    \caption{Deriving a photometric proxy for [OII] EW: We find a very strong correlation between [OII] EW and $U-R$ rest-frame colour.}
    \label{fig:colour_oII}
\end{figure}

As [OII] EW is derived using host galaxy spectroscopy, it would be advantageous to obtain a photometric alternative. One ideal candidate that has been shown to strongly correlate with Hubble residual is $U-R$ rest-frame colour (\citealp{2021Kelsey, Kelsey_2023}). In Figure \ref{fig:colour_oII}, we fit a linear model to our data and obtain the following relation:


\begin{equation}
    \mathrm{log(-[OII]\thinspace\text{EW}}) = -0.620\times(U-R) + 1.799.
\end{equation}

Both $U-R$ and the [OII] EW will depend on the stellar population, the amount of dust and its distribution. For example, star-forming regions, may be dustier than regions devoid of star formation. 

The bias correction relies on a model describing the residual scatter after standard light curve corrections are applied. Some of these models (e.g. P21) depend on host properties. The bias corrections could affect the correlations we measure. We find that using a dust-mass bias correction model \citep{Popovic_2021} does result in reducing the correlation between SN Ia brightness and host galaxy properties, but not entirely. \cite{Wiseman_2023} also discuss the potential improvements in having additional galaxy-age light curve correction parameters, where they find that sSFR is an ideal tracer for exploring this correlation with Hubble residuals.

Ultimately, the commonly used approach to correcting SN Ia light curves using host stellar mass does not fully remove the observed trends. Our analysis finds that the most effective strategy for improving the standardisation of SN Ia light curves involves both the `BBC4D' bias correction term and an additional parameter accounting for host sSFR, a tracer of stellar population age, instead of relying solely on stellar mass. We highlight the further impact that using [OII] EW can in correcting for host galaxy properties such as stellar mass or $U-R$. Importantly, this can be achieved with one spectroscopic feature.


\subsection{Future Work}

It would be interesting to increase our sample of SN Ia calibrators to cover all of the 37 galaxies in the SH0ES analysis. Additionally, we can extend our SN Ia calibrator sample to include TRGB \citep{Freedman_2021}, and SBF hosts \citep{Jensen_2021}. This would help provide more detail on the variation of SNe Ia in different galactic environments and perhaps a better understanding of the origin of differences in $H_{0}$ for different calibrators.

In the coming years, surveys such as TiDES \citep{2019Swann} will use 4MOST and will be able to observe and obtain spectra for tens of thousands of SN Ia host galaxies. For each galaxy, we can obtain a measurement of [OII] EW and a significantly larger sample size will help drive down statistical uncertainties. 

Additionally, the size of the OzDES sample allows one to explore how the relation changes with redshift. This can be probed with even greater statistical significance with TiDES. Redshift evolution could bias the value of cosmological parameters that are determined from SNe Ia.

The additional benefit of using the [OII] emission line, is that it can be observed up to $z\sim1$, unlike the stronger Balmer lines. However, this may bias the sample towards galaxies which are brighter, and have a higher sSFR. With increasing redshift, it becomes increasingly difficult to measure the [OII] EW. Ultimately, our work highlights the value of [OII] EW as an effective alternative to stellar mass in standardising SN Ia luminosities.

\section{Conclusion}
The SN Ia standardisable candle forms an important component of the astronomical distance ladder. However, correlations between Hubble residuals and the properties of the SN Ia hosts impact their use in measuring distances.
Using [OII] EW in our sample of 707 OzDES host galaxies, we obtain a correction that is more significant than the commonly used mass step correction. We then apply our [OII] EW correction to a sample of 20 SN Ia calibrator galaxies observed using WiFeS, and calibrate the SN Ia absolute magnitude. Applying this result to the Pantheon+ analysis, we find $H_{0}$ values ranging between 73.04$-$73.51$\hubble$, depending on our fitting approach in deriving the [OII] EW correction. The change in the value of $H_{0}$ is negligible when using the [OII] EW in place of or in addition to the mass of host galaxies in adjusting SN Ia luminosities. The tension between nearby and distant measurements of $H_{0}$ remains.


\section*{Acknowledgements}
We thank an anonymous referee for insightful feedback which improved the quality of the paper.
MD developed the paper with help from JM, CL, ENT, CF, and ARD who advised on the method, statistical anaylsis, assisted with observations, and provided feedback on the manuscript. LG and DS provided comments as internal reviewers, while valuable comments and assistance with the DES 5YR analysis were provided by TMD, AM, LK, JL, PW, MV, PS, BN AND BET. The remaining authors have made contributions to the paper through the DES and OzDES collaborations, which include data collection, analysis, pipeline development, validation tests, and promoting the science.

The author Mitchell Dixon would like to acknowledge support through an Australian Government Research Training Program Scholarship. This research was supported by the Centre of Excellence for Dark Matter Particle Physics (CDM; project number CE200100008) and the Australian Research Council Centre of Excellence for Gravitational Wave Discovery (OzGrav; project number CE170100004). This project/ publication was made possible through the support of a grant from the John Templeton Foundation. The authors gratefully acknowledge this grant ID 61807, Two Standard Models Meet. The opinions expressed in this publication are those of the author(s) and do not necessarily reflect the views of the John Templeton Foundation.

Funding for the DES Projects has been provided by the U.S. Department of Energy, the U.S. National Science Foundation, the Ministry of Science and Education of Spain, 
the Science and Technology Facilities Council of the United Kingdom, the Higher Education Funding Council for England, the National Center for Supercomputing 
Applications at the University of Illinois at Urbana-Champaign, the Kavli Institute of Cosmological Physics at the University of Chicago, 
the Center for Cosmology and Astro-Particle Physics at the Ohio State University,
the Mitchell Institute for Fundamental Physics and Astronomy at Texas A\&M University, Financiadora de Estudos e Projetos, 
Funda{\c c}{\~a}o Carlos Chagas Filho de Amparo {\`a} Pesquisa do Estado do Rio de Janeiro, Conselho Nacional de Desenvolvimento Cient{\'i}fico e Tecnol{\'o}gico and 
the Minist{\'e}rio da Ci{\^e}ncia, Tecnologia e Inova{\c c}{\~a}o, the Deutsche Forschungsgemeinschaft and the Collaborating Institutions in the Dark Energy Survey. 

The Collaborating Institutions are Argonne National Laboratory, the University of California at Santa Cruz, the University of Cambridge, Centro de Investigaciones Energ{\'e}ticas, 
Medioambientales y Tecnol{\'o}gicas-Madrid, the University of Chicago, University College London, the DES-Brazil Consortium, the University of Edinburgh, 
the Eidgen{\"o}ssische Technische Hochschule (ETH) Z{\"u}rich, 
Fermi National Accelerator Laboratory, the University of Illinois at Urbana-Champaign, the Institut de Ci{\`e}ncies de l'Espai (IEEC/CSIC), 
the Institut de F{\'i}sica d'Altes Energies, Lawrence Berkeley National Laboratory, the Ludwig-Maximilians Universit{\"a}t M{\"u}nchen and the associated Excellence Cluster Universe, 
the University of Michigan, NSF's NOIRLab, the University of Nottingham, The Ohio State University, the University of Pennsylvania, the University of Portsmouth, 
SLAC National Accelerator Laboratory, Stanford University, the University of Sussex, Texas A\&M University, and the OzDES Membership Consortium.

L.G. acknowledges financial support from the Spanish Ministerio de Ciencia e Innovaci\'on (MCIN) and the Agencia Estatal de Investigaci\'on (AEI) 10.13039/501100011033 under the PID2020-115253GA-I00 HOSTFLOWS project, from Centro Superior de Investigaciones Cient\'ificas (CSIC) under the PIE project 20215AT016 and the program Unidad de Excelencia Mar\'ia de Maeztu CEX2020-001058-M, and from the Departament de Recerca i Universitats de la Generalitat de Catalunya through the 2021-SGR-01270 grant.

AM is supported by the ARC Discovery Early Career Researcher Award (DECRA) project number DE230100055.

L.K. thanks the UKRI Future Leaders Fellowship for support through the grant MR/T01881X/1.

Based in part on observations at Cerro Tololo Inter-American Observatory at NSF's NOIRLab (NOIRLab Prop. ID 2012B-0001; PI: J. Frieman), which is managed by the Association of Universities for Research in Astronomy (AURA) under a cooperative agreement with the National Science Foundation.

The DES data management system is supported by the National Science Foundation under Grant Numbers AST-1138766 and AST-1536171.
The DES participants from Spanish institutions are partially supported by MICINN under grants ESP2017-89838, PGC2018-094773, PGC2018-102021, SEV-2016-0588, SEV-2016-0597, and MDM-2015-0509, some of which include ERDF funds from the European Union. IFAE is partially funded by the CERCA program of the Generalitat de Catalunya.
Research leading to these results has received funding from the European Research
Council under the European Union's Seventh Framework Program (FP7/2007-2013) including ERC grant agreements 240672, 291329, and 306478.
We  acknowledge support from the Brazilian Instituto Nacional de Ci\^encia
e Tecnologia (INCT) do e-Universo (CNPq grant 465376/2014-2).

This manuscript has been authored by Fermi Research Alliance, LLC under Contract No. DE-AC02-07CH11359 with the U.S. Department of Energy, Office of Science, Office of High Energy Physics.

Based on data acquired at the Anglo-Australian Telescope, under program A/2013B/012, and the 2.3-m Telescope across observing runs from 2020$-$2022.
We acknowledge the traditional custodians of the land on which the AAT stands, the Gamilaraay people, and pay our respects to elders past and present.

\section*{Data Availability}
The DES-SN photometric SN Ia catalogue will be made available as part of the DES5YR SN cosmology analysis at https://des.ncsa.illinois.edu/releases/sn.
The OzDES-DR2 spectra used in this paper can be publically accessed at: https://docs.datacentral.org.au/ozdes/overview/dr2.


\typeout{}
\bibliographystyle{mnras}
\bibliography{ref} 




\newpage
\appendix


\section{Author Affiliations}
$^{1}$ Centre for Astrophysics \& Supercomputing, Swinburne University of Technology, Victoria 3122, Australia\\
$^{2}$ Centre for Gravitational Astrophysics, College of Science, The Australian National University, ACT 2601, Australia\\
$^{3}$ The Research School of Astronomy and Astrophysics, Australian National University, ACT 2601, Australia\\
$^{4}$ Institut d'Estudis Espacials de Catalunya (IEEC), 08034 Barcelona, Spain\\
$^{5}$ Institute of Space Sciences (ICE, CSIC),  Campus UAB, Carrer de Can Magrans, s/n,  08193 Barcelona, Spain\\
$^{6}$ Department of Physics, Duke University Durham, NC 27708, USA\\
$^{7}$ School of Mathematics and Physics, University of Queensland,  Brisbane, QLD 4072, Australia\\
$^{8}$ Institute of Cosmology and Gravitation, University of Portsmouth, Portsmouth, PO1 3FX, UK\\
$^{9}$ School of Physics and Astronomy, University of Southampton,  Southampton, SO17 1BJ, UK\\
$^{10}$ Department of Physics and Astronomy, University of Pennsylvania, Philadelphia, PA 19104, USA\\
$^{11}$ Department of Physics \& Astronomy, University College London, Gower Street, London, WC1E 6BT, UK\\
$^{12}$ Laborat\'orio Interinstitucional de e-Astronomia - LIneA, Rua Gal. Jos\'e Cristino 77, Rio de Janeiro, RJ - 20921-400, Brazil\\
$^{13}$ Fermi National Accelerator Laboratory, P. O. Box 500, Batavia, IL 60510, USA\\
$^{14}$ Department of Physics, University of Michigan, Ann Arbor, MI 48109, USA\\
$^{15}$ University Observatory, Faculty of Physics, Ludwig-Maximilians-Universit\"at, Scheinerstr. 1, 81679 Munich, Germany\\
$^{16}$ Kavli Institute for Particle Astrophysics \& Cosmology, P. O. Box 2450, Stanford University, Stanford, CA 94305, USA\\
$^{17}$ SLAC National Accelerator Laboratory, Menlo Park, CA 94025, USA\\
$^{18}$ Instituto de Astrofisica de Canarias, E-38205 La Laguna, Tenerife, Spain\\
$^{19}$ Institut de F\'{\i}sica d'Altes Energies (IFAE), The Barcelona Institute of Science and Technology, Campus UAB, 08193 Bellaterra (Barcelona) Spain\\
$^{20}$ Jodrell Bank Center for Astrophysics, School of Physics and Astronomy, University of Manchester, Oxford Road, Manchester, M13 9PL, UK\\
$^{21}$ University of Nottingham, School of Physics and Astronomy, Nottingham NG7 2RD, UK\\
$^{22}$ Hamburger Sternwarte, Universit\"{a}t Hamburg, Gojenbergsweg 112, 21029 Hamburg, Germany\\
$^{23}$ Jet Propulsion Laboratory, California Institute of Technology, 4800 Oak Grove Dr., Pasadena, CA 91109, USA\\
$^{24}$ Institute of Theoretical Astrophysics, University of Oslo. P.O. Box 1029 Blindern, NO-0315 Oslo, Norway\\
$^{25}$ Kavli Institute for Cosmological Physics, University of Chicago, Chicago, IL 60637, USA\\
$^{26}$ Instituto de Fisica Teorica UAM/CSIC, Universidad Autonoma de Madrid, 28049 Madrid, Spain\\
$^{27}$ Center for Astrophysical Surveys, National Center for Supercomputing Applications, 1205 West Clark St., Urbana, IL 61801, USA\\
$^{28}$ Department of Astronomy, University of Illinois at Urbana-Champaign, 1002 W. Green Street, Urbana, IL 61801, USA\\
$^{29}$ Santa Cruz Institute for Particle Physics, Santa Cruz, CA 95064, USA\\
$^{30}$ Center for Cosmology and Astro-Particle Physics, The Ohio State University, Columbus, OH 43210, USA\\
$^{31}$ Department of Physics, The Ohio State University, Columbus, OH 43210, USA\\
$^{32}$ Center for Astrophysics $\vert$ Harvard \& Smithsonian, 60 Garden Street, Cambridge, MA 02138, USA\\
$^{33}$ Australian Astronomical Optics, Macquarie University, North Ryde, NSW 2113, Australia\\
$^{34}$ Lowell Observatory, 1400 Mars Hill Rd, Flagstaff, AZ 86001, USA\\
$^{35}$ Departamento de F\'isica Matem\'atica, Instituto de F\'isica, Universidade de S\~ao Paulo, CP 66318, S\~ao Paulo, SP, 05314-970, Brazil\\
$^{36}$ George P. and Cynthia Woods Mitchell Institute for Fundamental Physics and Astronomy, and Department of Physics and Astronomy, Texas A\&M University, College Station, TX 77843,  USA\\
$^{37}$ LPSC Grenoble - 53, Avenue des Martyrs 38026 Grenoble, France\\
$^{38}$ Instituci\'o Catalana de Recerca i Estudis Avan\c{c}ats, E-08010 Barcelona, Spain\\
$^{39}$ Department of Astrophysical Sciences, Princeton University, Peyton Hall, Princeton, NJ 08544, USA\\
$^{40}$ Department of Physics, University of Surrey, Guilford, Surrey, UK\\
$^{41}$ Observat\'orio Nacional, Rua Gal. Jos\'e Cristino 77, Rio de Janeiro, RJ - 20921-400, Brazil\\
$^{42}$ Department of Physics, Carnegie Mellon University, Pittsburgh, Pennsylvania 15312, USA\\
$^{43}$ Department of Physics, Northeastern University, Boston, MA 02115, USA\\
$^{44}$ Centro de Investigaciones Energ\'eticas, Medioambientales y Tecnol\'ogicas (CIEMAT), Madrid, Spain\\
$^{45}$ Instituto de F\'isica Gleb Wataghin, Universidade Estadual de Campinas, 13083-859, Campinas, SP, Brazil\\
$^{46}$ Computer Science and Mathematics Division, Oak Ridge National Laboratory, Oak Ridge, TN 37831\\
$^{47}$ Argonne National Laboratory, 9700 S Cass Ave, Lemont, IL 60439, USA\\
$^{48}$ Cerro Tololo Inter-American Observatory, NSF's National Optical-Infrared Astronomy Research Laboratory, Casilla 603, La Serena, Chile\\
$^{49}$ Department of Astronomy, University of California, Berkeley,  501 Campbell Hall, Berkeley, CA 94720, USA\\
$^{50}$ Lawrence Berkeley National Laboratory, 1 Cyclotron Road, Berkeley, CA 94720, USA\\
$^{51}$ INAF-Osservatorio Astronomico di Trieste, Via G. B. Tiepolo 11, 34143 Trieste, Italy


\bsp	
\label{lastpage}
\end{document}